\documentclass[11pt]{article}
\usepackage[utf8]{inputenc}
\usepackage{caption}
\usepackage{braket,slashed,bm}
\usepackage{jheppub}
\usepackage{simplewick}
\usepackage{array,multirow}
\usepackage{amsmath} \DeclareMathAlphabet{\mathpzc}{OT1}{pzc}{m}{it}
\usepackage[normalem]{ulem}
\usepackage{calligra}
\usepackage{floatrow}
\DeclareMathAlphabet{\mathpzc}{OT1}{pzc}{m}{it}
\usepackage{mathrsfs}
\usepackage{tikz}
\usepackage{verbatim}
\usepackage{cleveref}

\crefname{section}{Sec.}{Secs.}
\crefname{table}{Tab.}{Tabs.}
\crefname{figure}{Fig.}{Figs.}
\crefname{equation}{Eq.}{Eqs.}
\crefname{appendix}{Appendix}{Appendix}

\usepackage{pgf}

\usepackage{subcaption}
\usepackage{lscape}
\usepackage{hyperref}
\usepackage{graphicx}
\usepackage{booktabs}
\newfloatcommand{capbtabbox}{table}[][\FBwidth]

\usepackage{url}

\newcommand{\amc}{{\sc MadGraph5\textunderscore}a{\sc MC@NLO}}
\newcommand{\frules}{{\sc Feyn\-Rules }}

\newcommand{\ra}[1]{\renewcommand{\arraystretch}{#1}}  

\graphicspath{{./Figures/}}

\newcommand{\mpt}{{\;/\!\!\!\! {P}_T}} 
\newcommand{\mptvec}{{\;/\!\!\!\! \vec{P}_T}}
\newcommand{\met}{{\:/\!\!\!\! E_T}} 
\def\beq{\begin{equation}}
\def\eeq{\end{equation}}
\def\bea{\begin{eqnarray}}
\def\eea{\end{eqnarray}}

\pgfkeys{/pgf/number format/.cd ,precision=2,sci generic={exponent={\times 10^{#1}}}}

\allowdisplaybreaks

\begin{document}

\title{Uncovering doubly charged scalars with dominant three-body decays using machine learning
}

\affiliation[a]{Center for AI and Natural Sciences, KIAS, Seoul 02455, Republic of Korea}

\affiliation[b]{Department of Physics, Chungbuk National University, Cheongju, Chungbuk 28644, Korea}

\affiliation[c]{Institut f\"{u}r Theoretische Physik und Astrophysik, Uni W\"{u}rzburg, Emil-Hilb-Weg 22, D-97074 W\"{u}rzburg, Germany}

\affiliation[d]{School of Physics, KIAS, Seoul 02455, Republic of Korea}

\affiliation[e]{Center for Theoretical Physics of the Universe,
Institute for Basic Science, Daejeon 34126, Republic of Korea}

\author[a]{Thomas Flacke,}
\emailAdd{flacke@kias.re.kr}
\author[b,d,e]{Jeong Han Kim,}
\emailAdd{jeonghan.kim@cbu.ac.kr}

\author[c]{Manuel Kunkel,}
\emailAdd{manuel.kunkel@physik.uni-wuerzburg.de}

\author[a,d]{Pyungwon Ko,}
\emailAdd{pko@kias.re.kr}

\author[b]{Jun Seung Pi,}
\emailAdd{junseung.pi@cbu.ac.kr}

\author[c]{Werner Porod,}
\emailAdd{porod@physik.uni-wuerzburg.de}

\author[c]{Leonard Schwarze}
\emailAdd{leonard.schwarze@stud-mail.uni-wuerzburg.de}

\abstract{We propose a deep learning-based search strategy for pair production of doubly charged scalars undergoing three-body decays to $W^+ t\bar b$ in the same-sign lepton plus multi-jet final state. 
This process is motivated by composite Higgs models with an underlying fermionic UV theory. We demonstrate that for such busy final states, jet image classification with convolutional neural networks outperforms standard fully connected networks acting on reconstructed kinematic variables. We  derive the expected discovery reach and exclusion limit at the high-luminosity LHC.
}

\preprint{KIAS-A23009}

\maketitle

\section{Introduction}
\label{sec:intro}

The first two runs of the Large Hadron Collider (LHC) have shown that the Standard Model
(SM) of particle physics continues to be an excellent description of Nature. Many of its features
have been tested to an extremely high accuracy by collider experiments over the last four
decades. The discovery of the Higgs boson, a crucial prediction of the SM, by the ATLAS
and CMS collaborations at the LHC in 2012 \cite{ATLAS:2012yve,CMS:2012qbp} has strengthened confidence in the ability
of the SM to accurately describe subatomic physics. Despite its enormous success, several
important features of Nature are unexplained, thus motivating the search for physics beyond
the SM. Among them are: (i) The origin of the electroweak sector: why does the Higgs sector
have the SM-form, in particular why is the squared mass parameter $\mu^2$ negative? (ii)~Are there
any additional scalars besides the observed Higgs boson? (iii) Is the observed Higgs boson a
fundamental scalar or does it have an internal structure?

An interesting path for explaining the origin of electroweak symmetry breaking is given in composite Higgs models
\cite{Kaplan:1983fs,Kaplan:1983sm,Contino:2003ve,Agashe:2004rs}. Composite extensions of the SM are very
well motivated due to their similarity with QCD as an example of a strongly coupled theory yielding bound states at low scales which is realised in nature. In many composite Higgs models, the Higgs boson is a pseudo Nambu-Goldstone boson (pNGB),
which explains its relatively small mass compared to other states predicted in these models.
A new gauge symmetry ``hypercolour'' is postulated with new ``hyperquarks’'
charged under this symmetry. The corresponding coupling constant, although small at high energies, becomes
large at a scale $\Lambda_\mathrm{HC} \sim O($TeV$)$ where a condensate is formed.
It is further assumed that a global
symmetry  group $G$ is present above $\Lambda_\mathrm{HC}$ which gets broken to a subgroup $H$ by the condensate, yielding at least four
Nambu-Goldstone bosons which form -- within the SM description -- the corresponding Higgs doublet. In a second step the
couplings of the SM gauge fields, quarks, and
leptons are taken into account which amount to an explicit breaking of $G$. In this way a potential for the
pNGBs is radiatively generated, which then leads to electroweak symmetry breaking. The hierarchical
Yukawa couplings can be  explained via a mechanism known as partial compositeness, in which
the SM fermions mix with composite states \cite{Kaplan:1991dc}.

The so-called ``minimal'' composite Higgs model \cite{Agashe:2004rs,Agashe:2005dk} is based on the symmetry breaking pattern SO(5)/SO(4), where the only four pNGBs precisely match the four degrees of freedom of the SM Higgs doublet field. The phenomenology of this model has been widely explored, and we refer the interested reader to the excellent reviews \cite{Contino:2010rs,Bellazzini:2014yua,Panico:2015jxa}. The minimal model, and variations thereof, can be considered as a useful template for understanding the phenomenology of a composite pNGB Higgs. However, models in this class are not UV complete. 
From the point of view of an underlying fermionic gauge theory, \`a la QCD, the symmetry breaking pattern SO(5)/SO(4) cannot be obtained as the global symmetry group of the underlying fermions is unitary \cite{Cacciapaglia:2014uja,Cacciapaglia:2020kgq}. 
The minimal breaking patterns for a composite Higgs which arises from models with underlying fermions are SU(4)/Sp(4), SU(5)/SO(5) or SU(4)$\times$SU(4)/SU(4) \cite{Dugan:1984hq,Galloway:2010bp,Ferretti:2013kya,Cacciapaglia:2014uja,Ma:2015gra}. All of these models contain additional pNGBs besides the Higgs doublet. 
Several aspects of their LHC phenomenology have been investigated in the literature, see for example \cite{Belyaev:2016ftv,Agugliaro:2018vsu,Banerjee:2022izw,Banerjee:2022xmu,Cacciapaglia:2022bax} and refs.~therein. In particular, one can obtain strong bounds on the masses of these additional pNGBs in final states containing photons and multi-lepton final states, where the latter stem from decays of vector bosons \cite{Banerjee:2022xmu,Cacciapaglia:2022bax}. While these final states are clean at a hadron collider like the LHC, the corresponding branching ratios can be rather small. In contrast, the branching ratios for decays into quarks, in particular top and bottom quarks, are usually significantly larger but the corresponding signals suffer from large backgrounds. As a consequence, the decays into quarks are less strongly constrained. Busy final states comprised of multiple jets and leptons face numerous challenges such as multiple jet reconstructions at the cost of signal efficiency, and the combinatorial problem to correctly pair them to resonances.
This motivates the development of new methods, and in this article we address this challenge by using machine learning (ML) techniques.

Deep neural networks (DNN) have proven to be powerful for image recognition~\cite{LeCun:2012,Krizhevsky:2012}, and have become a ubiquitous tool in high energy physics to search for new particles~\cite{Bhimji:2017qvb, Andrews:2019wng, Huang:2022rne, Kim:2019wns, Kim:2022miv, Diefenbacher:2019ezd,Kim:2023wuk}.
Hadronic activity can be visualised in discretised calorimeter grids -- so-called jet images \cite{deOliveira:2015xxd, Cogan:2014oua, Kasieczka:2019dbj} -- which contain information about the direction and the energy of particles. Jet images also contain information on the difference in colour flow between the signal and background events. We note for completeness, that ML techniques have also been proposed for generic BSM searches, see e.g.~\cite{Hajer:2018kqm,Farina:2018fyg,Heimel:2018mkt,Blance:2019ibf,CrispimRomao:2020ucc,Bortolato:2021zic,Hallin:2021wme,Jawahar:2021vyu,Ngairangbam:2021yma,Karagiorgi:2021ngt}, 
but they all come with certain theory assumptions and limitations as has been critically accessed in \cite{Finke:2021sdf}.

In this article we focus on pair production of doubly charged scalars which subsequently decay via the three-body decay $S^{++}\rightarrow W^+ t \bar{b}$ and the conjugate process. This signature is theoretically well motivated in underlying models of a composite Higgs and has been shown to be only weakly constrained by current LHC searches \cite{Cacciapaglia:2022bax}. We focus on the leptonic decay of either the $W^+ t$ or the $W^- \bar{t}$ system which yields a same-sign di-lepton signature. Moreover, the signal involves a lot of hadronic activity from the decays of the other particles in this channel. We target this hadronic activity by a combination of jet images and kinematic variables of reconstructed objects and we show that deep neural networks can provide powerful discrimination between signal and background.

This paper is organised as follows: In \cref{sec:model} we summarise the theoretical background underlying the process we study, followed by details on the event generation and preselection cuts in \cref{sec:eventgen}. Section~\ref{sec:input} then describes the data representation and the machine learning setup. We present our results in \cref{sec:results}, and draw our conclusions and present an outlook in \cref{sec:conclusion}. We provide additional information in two technical appendices. \cref{sec:appen} gives details about the neural network architectures and comparisons with networks not used in the main text. Finally, in \cref{sec:appeval} we provide additional information about the evaluation techniques and comparisons between different training data sets.

\section{Extended scalar sectors in composite Higgs models} 
\label{sec:model}

The cosets of composite Higgs models with a fermionic UV completion contain additional pNGBs besides the Higgs doublet. 
In case of SU(4)/Sp(4) \cite{Galloway:2010bp,Cacciapaglia:2014uja,Ferretti:2016upr} the additional pNGB is just a gauge singlet. For SU(5)/SO(5) \cite{Dugan:1984hq,Ferretti:2016upr} there is a gauge singlet and three  SU(2)$_L$ triplets with hypercharges $Y=0,\pm 1$. For SU(4)$\times$SU(4)/SU(4) \cite{Ma:2015gra,Ferretti:2016upr} there are a SU(2)$_L$ triplet with $Y=0$, an additional SU(2)$_L$ doublet with $Y= -1$ and three SU(2)$_L$ singlets with $Y=0,\pm 1$. 
The decay modes of these additional pNGBs depend
on the details of the underlying model but fall into three broad classes \cite{Ferretti:2016upr,Agugliaro:2018vsu}:
\begin{itemize}
	\item decays into electroweak gauge bosons via Wess-Zumino-Witten (WZW) terms \cite{Wess:1971yu,Witten:1983tw},
	\item decays into an electroweak gauge boson and another pNGB. These dominate over the decays via WZW terms if the mass splitting between the pNGBs is larger than a few GeV \cite{Banerjee:2022xmu,Cacciapaglia:2022bax}, or
	\item decays into two quarks in case of models with partial compositeness. Here, one expects the decays into top and bottom quarks as these have the largest masses. These usually dominate for mass splitting between the pNGBs of up to 50 GeV \cite{Banerjee:2022xmu,Cacciapaglia:2022bax}.
\end{itemize}
It has been shown in \cite{Banerjee:2022xmu,Cacciapaglia:2022bax} that existing LHC data put strong bounds on the masses of the electroweak pNGBs of up to 700 GeV if the decays into electroweak gauge bosons dominate. The bounds are considerably weaker in the case that the decays into quarks dominate. 
In the following, we investigate to which extent these bounds can be improved by employing ML techniques. 
For concreteness, we focus on the SU(5)/SO(5) coset.
After electroweak symmetry breaking the triplets can be placed into multiplets of the remaining custodial SU(2): a singlet $\eta_1$, a triplet $\eta_3$, and a quintuplet $\eta_5$. The latter contains a doubly charged member $S^{++}$ and a singly charged $S^+$, which will serve as reference states in the following analysis.
One expects that the members of a multiplet have approximately the same mass~\cite{Ferretti:2016upr,Agugliaro:2018vsu}, so we take $m_{S^{++}} = m_{S^+}$.

\begin{figure*}[t!]
	\begin{center}
		\includegraphics[width=0.4\textwidth,clip]{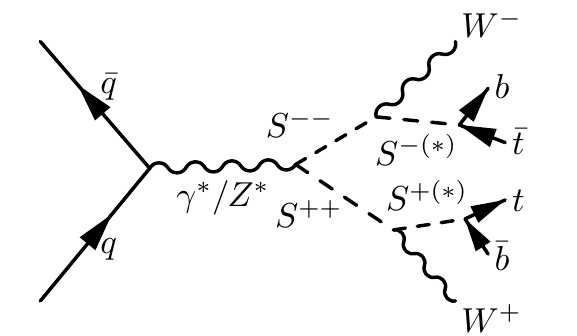} 
		\includegraphics[width=0.4\textwidth,clip]{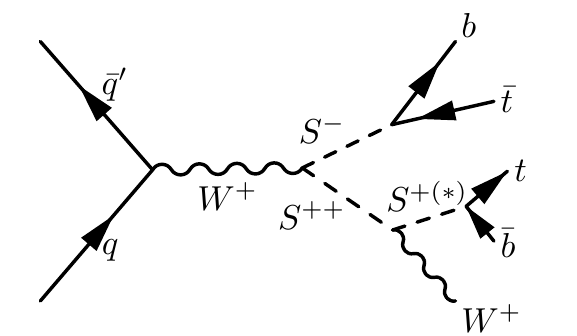} 
		\caption{The pair production of doubly charged scalars (left), and mixed pNGB productions (right) via Drell-Yan processes.
			\label{fig:feynman}}
	\end{center}
\end{figure*}

At the LHC the electroweak pNGBs are produced in Drell-Yan processes as shown in \cref{fig:feynman}. One usually assumes
that the $S^{++}$ either decays into two $W^+$ or into two leptons which is motivated by seesaw type II models \cite{Schechter:1980gr}. 
While these decay channels have been searched for \cite{ATLAS:2021jol,Leban:2022pdz,CMS:2017pet}, in composite Higgs models they can be subdominant to decays into quarks, which are the focus of this work.
The singly charged scalar decays according to $S^+ \to t \bar{b}$.
Although the $S^{++}$ does not have a direct coupling to quarks, it decays via an off-shell $S^+$ (see also \cref{fig:feynman})
\begin{align}
S^{++} \to W^+ S^{+*} \to W^+ t \bar{b} \,.
\end{align}
In case of rather large mass splitting of 40 or more GeV among the pNGBs, the decay into an off-shell $W$-boson and an on-shell $S^+$ becomes important~\cite{Cacciapaglia:2022bax}. 
At the LHC, these scalars can either be produced as decay products of heavier bound states like top partners, see e.g.~\cite{Banerjee:2022xmu,Kim:2018mks, Kim:2019oyh}, or directly via Drell-Yan processes.
In what follows, we concentrate on the second option as it might be that bound states other than the pNGBs are too heavy to be produced at the LHC. 
In the model-specific examples below, the two scalars $S^{++}$ and $S^{+}$ are assumed to have the same quantum numbers as in the quintuplet in the SU(5)/SO(5) coset~\cite{Agugliaro:2018vsu,Cacciapaglia:2022bax}.

\section{Event generation and preselection cuts}
\label{sec:eventgen}

The electroweak pNGBs described in \cref{sec:model} open up new and under-explored final states at the LHC. 
Typical signatures comprise of multiple electroweak bosons alongside with top and bottom quarks, giving rise to very busy final states.
However, what remains to be addressed is how we will identify such a considerable amount of jets and leptons at the cost of signal efficiency, and whether we can distinguish these signatures from SM backgrounds.
To account for these challenges, we need to develop new deep learning-based strategies beyond traditional cut-and-count methods to increase the potential discovery reach of these new particles.

As a first step towards this path, we consider the Drell-Yan pair production of doubly charged scalars
\begin{equation} 
q \bar{q} \rightarrow S^{++} S^{--}  \rightarrow (W^+ t \bar{b}) \; (W^- \bar{t} b)
\rightarrow (W^+ W^+ b \bar{b}) \; (W^- W^- b \bar{b})\;,
\label{eq:prod1}
\end{equation} 
where the three-body decays are mediated by an off-shell $S^{\pm*}$ as shown in \cref{fig:feynman} (left). In the scenario where the pNGBs are allowed to couple to SM fermions, the pNGBs dominantly decay into third generation quarks because the couplings with electroweak gauge bosons are one-loop suppressed. Once the top quarks decay, therefore, the final state consists of four $W$ bosons and four bottom quarks, where the same-sign $W$ bosons arise from the same decay chain.

\begin{figure*}[t!]
	\begin{center}
		\includegraphics[width=0.5\textwidth,clip]{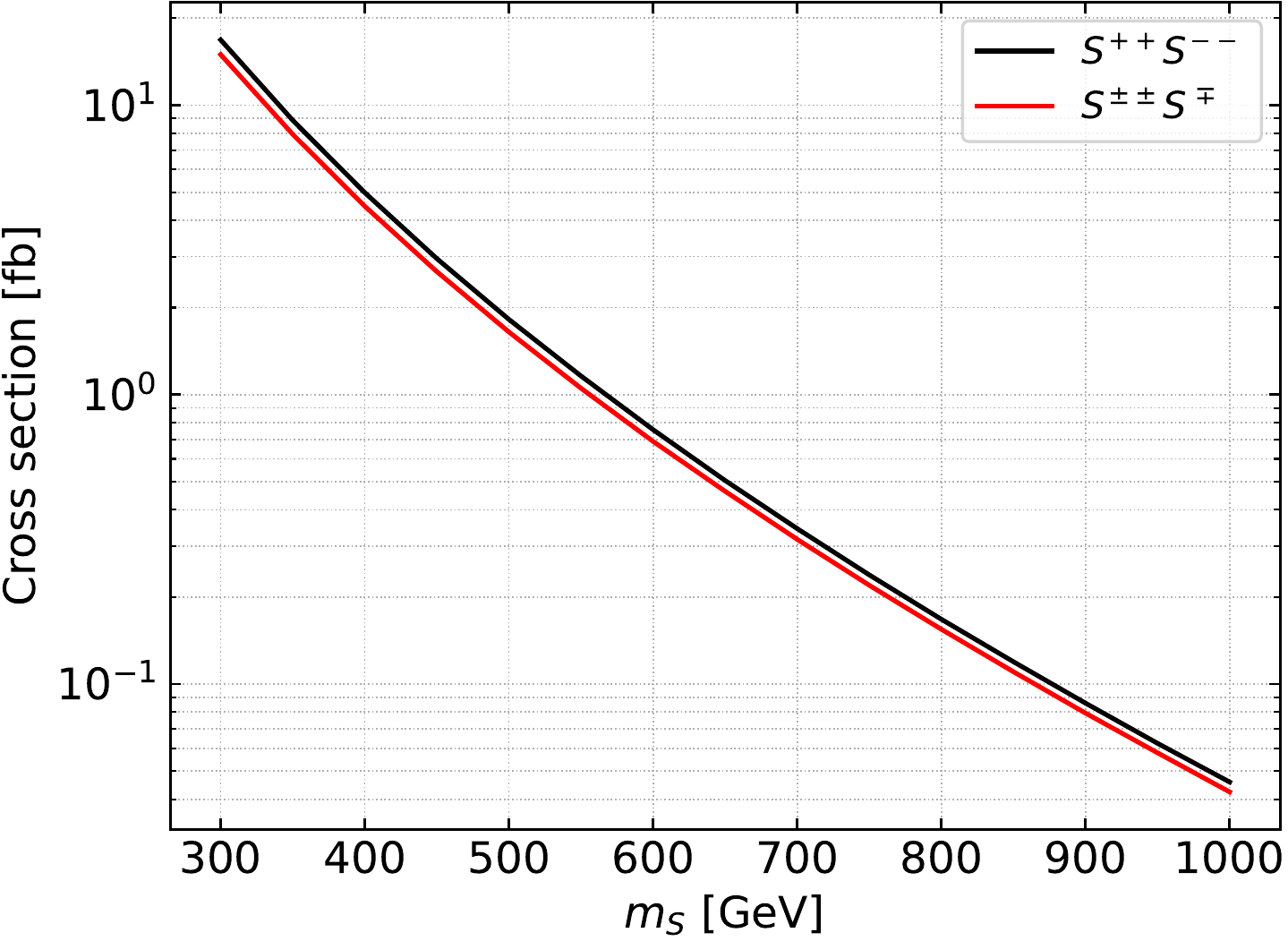} 
		\caption{
			Drell-Yan production cross sections of $S^{++}S^{--}$ (black) and $S^{\pm\pm}S^{\mp}$ (red) at $\sqrt s=14$~TeV.}\label{fig:Crosssection}
	\end{center}
\end{figure*}

In the presence of several other electroweak pNGBs in the same multiplet, however, this final state has similar characteristics to the mixed pNGB production shown in \cref{fig:feynman} (right)
\begin{equation} 
q \bar{q}' \rightarrow S^{\pm\pm} S^{\mp}  \rightarrow (W t b) \; (t b)
\rightarrow (W W b b) \; (W b b) \;,
\label{eq:prod2}
\end{equation} 
which also yields same-sign $W$ bosons, but in total one $W$ boson fewer than the process in \cref{eq:prod1}.
Figure~\ref{fig:Crosssection} shows that the production cross sections of both processes are of a similar order of magnitude, and their kinematic distributions are similar because they both result from Drell-Yan production. Thus, after a series of selection cuts, the mixed pNGB production will likely contribute to the signal as well. 

The relative weight of the production cross sections between the two processes depends on the specific realisation of the model. Here, to incorporate this dependence, we explore two scenarios: i) when only the doubly charged scalar production is present; ii) when both the doubly charged scalar and mixed pNGB productions are present, where their relative cross sections are taken from the $\eta_5$ multiplet described in \cref{sec:model}.

Given multiple $W$ bosons in the final states, there are a number of possible decay channels to explore. To focus on the channel with smaller SM backgrounds,
we consider the case where a pair of same-sign $W$ bosons decay leptonically, while the remaining $W$ bosons decay hadronically. This gives rise to the final state with two same-sign leptons, missing transverse momentum, four $b$-tagged jets, and several light-flavour jets.

For the numerical study of the signal processes in \cref{eq:prod1,eq:prod2}, we use the public \frules \cite{Alloul:2013bka} implementation of the eVLQ model~\cite{Banerjee:2022xmu,Flacke:eVLQ} which provides generic interactions of the pNGBs and vector-like quarks. 
We use this implementation to generate a UFO library \cite{Degrande:2011ua} which allows for event generation within the \amc{} \cite{Alwall:2014hca,Hirschi:2015iia} framework.
We generate the signal events for a centre-of-mass energy of $\sqrt{s} = 14$~TeV at the leading order (LO) in QCD with NNPDF2.3QED parton distribution functions \cite{Ball:2013hta} using dynamical renormalisation and factorisation scales. 
We apply a flat $K$-factor of 1.15 to the cross sections to account for NLO effects  \cite{Fuks:2013vua}.
In \cref{fig:Crosssection}, we show the production cross sections as a function of the mass of the pNGBs. In this analysis, we assume that all pNGBs are mass degenerate. 

We generate SM backgrounds for a centre-of-mass energy of $\sqrt{s} = 14$ TeV at the next-to-leading-order (NLO) in QCD (unless mentioned otherwise) using the same PDF and scale choices.
The irreducible $t\bar{t}t\bar{t}$ background is generated at LO accuracy of which the cross section is normalised to the NLO cross section of $1.17 \times 10^{-2}$~pb. 
The next most important backgrounds are $t\bar{t}h$ and $t\bar{t}V$,  where $V$ denotes an electroweak gauge boson $W$ or $Z$, whose NLO cross sections are $5.60 \times 10^{-1}$ pb and $1.60$ pb, respectively. A subdominant background is given by $t\bar{t}VV$ with NLO cross section of $1.89 \times 10^{-2}$ pb.
Other backgrounds, such as $VV$ and $VVV$, turn out to be negligibly small (after preselection cuts and after taking branching ratios into account), and we therefore we do not consider them further in the following analysis.

After generating hard scattering events, the events are passed to 
{\sc Pythia8}~\cite{Sjostrand:2014zea} for showering and hadronisation.
We use {\sc Delphes} 3.4.1 \cite{deFavereau:2013fsa} to include detector resolution effects based on modified ATLAS configurations \cite{Kim:2019wns}.
For jet reconstruction, we utilise the {\sc Fastjet} 3.3.1 \cite{Cacciari:2011ma} implementation with an anti-$k_T$ algorithm and a cone radius of $r = 0.4$. 
We benchmark a HL-LHC $b$-tagging efficiency from ATLAS report~\cite{CERN-LHCC-2017-021},
where we use a flat $b$-tag rate of $\epsilon_{b \rightarrow b} = 0.8$, and a mistag rate for a $c$-jet (light-flavor jet) being misidentified as a $b$-jet of $\epsilon_{c \rightarrow b} = 0.2$ ($\epsilon_{j \rightarrow b} = 0.01$).

We require isolated leptons to satisfy $p_T(\ell)/(p_T(\ell) + \sum_i p_{T,i})>0.7$ where $\sum_i p_{T,i}$ is the sum of the transverse momenta of nearby particles with $p_{T,i} > 0.5$~GeV and $\Delta R_{i\ell} < 0.3$. Isolated leptons are required to pass minimum cuts $p_T(\ell) > 20$~GeV and $|\eta(\ell)| < 2.5$. The charge misidentification of leptons is taken into account by \textsc{Delphes}.

Although it can be useful to fully identify final state objects to reconstruct heavy resonances, it is typically accompanied by a poor signal acceptance due to the preselection criteria described above. Besides, the combinatorial problem makes it difficult to correctly associate all jets to the particles they originate from, given the very busy final state.
To cope with this, we propose a more inclusive search by requiring at least three $b$-tagged jets ($N(b) \geq 3$) and at least three non-$b$-tagged jets ($N(j) \geq 3$).
Finally, events are required to have exactly two same-sign leptons, and to pass the minimum missing transverse momentum (defined as in Ref. \cite{Kim:2019wns}) $\mpt = | \mptvec | > 20 $ GeV, and $S_T > $ 400 GeV where $S_T$ is the scalar sum of the transverse momenta of the reconstructed jets and the two same-sign leptons. 

The signal and background cross sections after this baseline selection are summarised in \cref{tab:Basiccut}.
In this table, we take the reference pNGB mass for signal events to be $m_S = 400$ GeV.
The first and second columns show the list of signal and backgrounds and their corresponding efficiencies of the  preselection.
The third and fourth columns show the cross sections and number of events for the target integrated luminosity ${\cal L}$ = 3 ab$^{-1}$ of the HL-LHC.

\begin{table}[]
	\ra{1.15}
	\begin{tabular}{lccc}
		\toprule
		Process            & $\epsilon_\mathrm{Preselection}$ & Cross section [fb]   & Events at 3~ab$^{-1}$ \\ \midrule
		$S^{++} S^{--}$    & $9.87\times 10^{-3}$             & $4.90\times 10^{-2}$ & 147                  \\
		$S^{\pm\pm} S^\mp$ & $4.81\times 10^{-3}$             & $2.87\times 10^{-3}$ & 86                   \\
		$t\bar t V$        & $1.70\times 10^{-4}$             & $2.72\times 10^{-1}$ & 816                  \\
		$t\bar th$         & $3.75\times 10^{-4}$             & $2.10\times 10^{-1}$ & 629                  \\
		$t\bar tt\bar t$   & $1.63 \times 10^{-2}$            & $1.91\times 10^{-1}$ & 572                  \\
		$t\bar t VV$       & $1.74\times 10^{-3}$             & $3.29\times 10^{-2}$ & 98                   \\
		$VVV$              & $2.08\times 10^{-6}$             & $1.05\times 10^{-3}$ & 3                    \\ \bottomrule
	\end{tabular}
	\caption{Signal and background efficiencies and cross sections after the preselection cuts. For signal processes, we take the reference pNGB mass to be $m_S = 400$ GeV. }
	\label{tab:Basiccut}
\end{table}

\section{Input data and network architectures}
\label{sec:input}

The  preselection cuts described in \cref{sec:eventgen} provide a preparation for a more sophisticated machine learning analysis.
A deep learning-based analysis can efficiently exploit differences in differential distributions, which, as compared to traditional cut-and-count methods, increases the discrimination power in searches for new particles.
However, before coming to the description of the network architecture, we need to decide on the data representation and data pre-processing suitable for the individual architectures. Possibilities range from training a DNN on the lowest level detector data to using neural networks with human-engineered kinematic variables which are targeted to specific features of the signal event (like for example reconstructing invariant masses of intermediate states). For the analysis we propose here, we use as inputs kinematic information of conventional reconstructed objects (electrons, muons, jets, $b$-tagged jets) as well as a number of simple derived kinematic quantities (invariant masses of two object pairs, radial distance of object pairs, missing transverse energy, and total transverse energy) as is described in more detail in \cref{sec:kinematic variavles}. Additionally, we make use of image recognition techniques for which events are translated to ``jet images'' which are described in \cref{sec:jetimages}.

To distinguish the signal from the background events based on this data representation, we have tested several network architectures. For details on the various architectures and performance comparisons, we refer the interested reader to the extensive Appendix \ref{sec:appen}. In \cref{sec:MLaspects} we summarise the network architecture which we found to perform best. 
This turned out to be a marriage of two smaller networks: a convolutional neural network (CNN) used for discrimination of jet images, and a fully connected (FC) deep neural network which only works with kinematic variables. The outputs of both networks are then interfaced for a combined classification.

\subsection{Jet images}\label{sec:jetimages}

Jet images are based on the particle flow information in each event.
Following a similar procedure as described in Ref.~\cite{Huang:2022rne},
we divide the particle flow into charged and neutral particles. The charged particles include charged hadrons, while the neutral particles consist of neutral hadrons as well as non-isolated photons. Leptons are removed from both samples. 
We define the origin of the ($\eta$, $\phi$) coordinates to be the centre of the two same-sign isolated leptons.
All particles are mapped accordingly in the ($\eta$, $\phi$) plane. The jet images are discretised into $50 \times 50$ calorimeter grids within a region of $-2.5\le \eta \le 2.5$ and $-\pi \le \phi \le \pi$. The intensity of each pixel is given by the total transverse momentum of particles passing through the pixel.\footnote{This information is obtained after showering and hadronising at the {\sc Delphes} level. When summing transverse momenta of hadrons in each pixel, we only consider the hadrons with $p_T>0.7$~GeV.} The final jet images have a dimension of $(2 \times 50 \times 50)$ where 2 denotes the number of channels, charged and neutral particle images. In order for the neural networks to fully take into account the spatial correlation between images of final state particles, we project the two same-sign leptons into the discretised $50 \times 50$ calorimeter grids as well. Combined with the jet images, we have a set of images for visible particles whose data structure is represented by
\begin{align}  
\mathcal I_{CN\ell} = \big( 3 \times 50 \times 50  \big)  \; ,
\label{eq:3images}
\end{align}
where 3 denotes charged ($C$), neutral ($N$) particle images, and lepton ($\ell$) images.

\begin{figure}
	\centering
	\includegraphics[width=\textwidth]{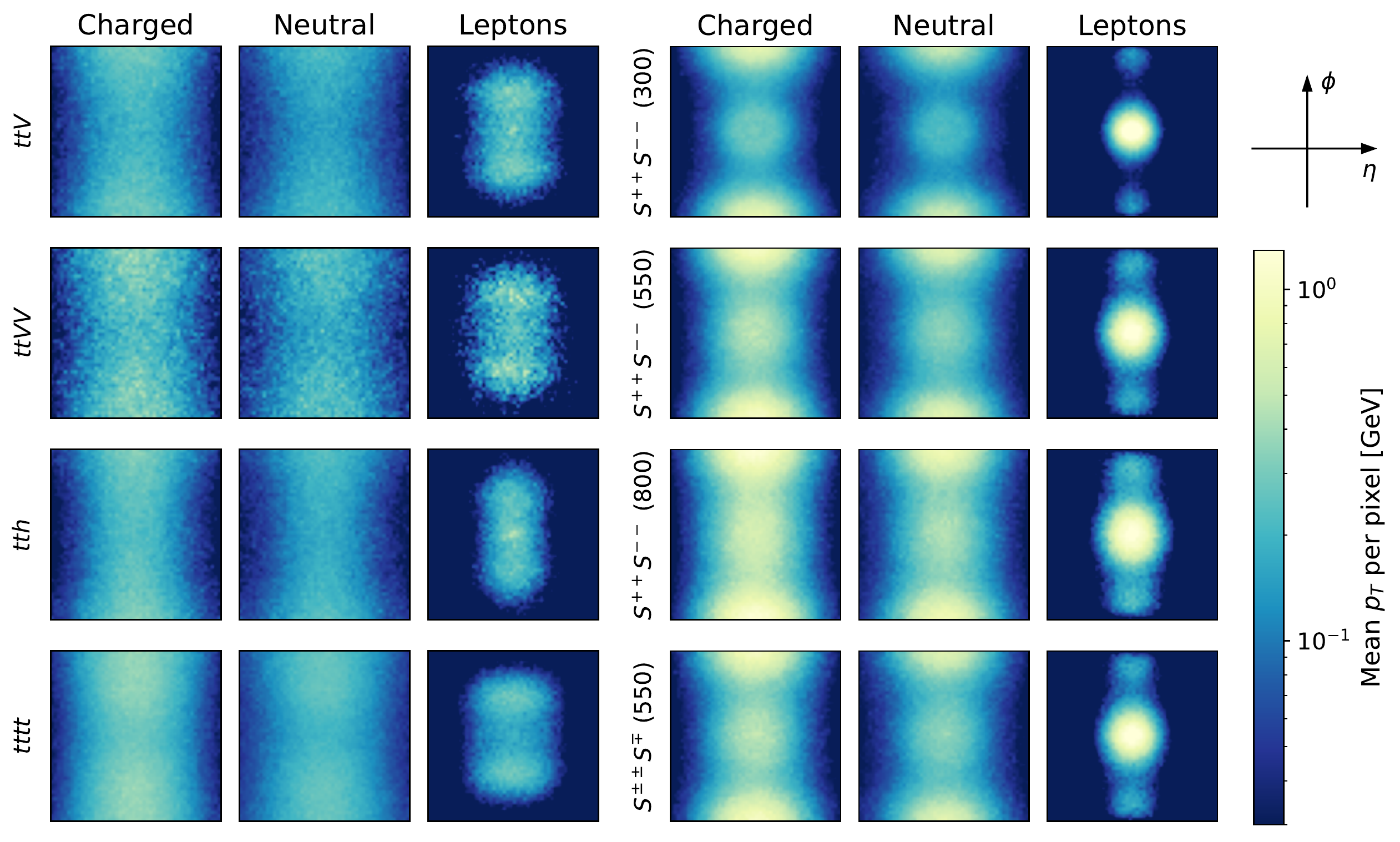}
	\caption{Distribution of jet and lepton images of background (left) and signal (right) processes, where the scalar mass is given in parentheses. The final state signatures are discretised into $50 \times 50$ calorimeter grids within a region of $-2.5\le \eta \le 2.5$ and $-\pi \le \phi \le \pi$. The origin of the coordinate system is the centre of the two same-sign leptons. The columns show charged and neutral components of the jets, as well as isolated leptons, respectively. The colourbar indicates the mean $p_T$ per pixel.}\label{fig:jetimages}
\end{figure}

In \cref{fig:jetimages} we show an overlay of the jet images of all simulated events in the background categories (left three columns) as well as the signal categories (right three columns), where the respective scalar mass is indicated in parentheses.
When comparing the jet images of signal and background processes, several differences are apparent: 
The signal images are generally ``brighter'' than those of the backgrounds, indicating a larger total $p_T$. The brightness increases with higher  mass of the scalars.
Furthermore, for the signal events the leptons are mostly distributed near the origin, while they are more broadly distributed for the backgrounds.
The reason is that in case of the backgrounds, the same-sign leptons originate from the decays of different particles. This leads to a larger angular separation compared to the signals, where both same-sign leptons come from the same $S^{++}$ or $S^{--}$.

One potential issue with using $\mathcal I_{CN\ell}$ is that
pileup effects could jeopardise the analysis, as the expected average number of pileup interactions $\langle \mu \rangle$ at the HL-LHC is $\mathcal{O}(200)$ collisions per bunch crossing \cite{ATLAS:2019xli}. We refer to Ref.~\cite{Kim:2019wns} for the semi-realistic examination of pileup effects, and several methods to mitigate the contamination.
Although charged particles can be potentially cleaned up from pileup by scrutinising the longitudinal vertex information \cite{Bertolini:2014bba}, neutral particles cannot be treated the same way. 
There are, however, on-going experimental works to mitigate the impact of neutral pileup collisions~\cite{CMS:2020ebo,Bertolini:2014bba}.
In addition to our analysis based on the $\mathcal I_{CN\ell}$, we therefore also perform a more conservative analysis based on the image data set excluding neutral particles whose data structure is represented by
\begin{align}  
\mathcal I_{C\ell} = \big( 2 \times 50 \times 50  \big)  \; .
\label{eq:2images}
\end{align}
The impact of excluding the neutral particle image will be discussed in \cref{sec:results}.

\subsection{Kinematic variables}
\label{sec:kinematic variavles}

After the basic selections described in \cref{sec:eventgen}, we consider two same-sign leptons, three leading $b$-tagged jets, and three leading non-$b$-tagged jets to construct kinematic variables. We do not utilise other inclusive jets in the analysis.
The most basic kinematic variables are four-momenta of the reconstructed objects.
From these, we can construct higher level kinematic variables which often serve as efficient learnable features for neural networks.
We consider the following set of kinematic variables
\begin{equation}
\mathcal K = \bigcup_{i \neq j} M_{ij} \cup \bigcup_{i \neq j }\Delta R_{ij} 
\cup \bigcup_{i } p_{Ti} \cup \{ \met, S_T \} \;  ,\label{eq:5kin}
\end{equation}
where $M_{ij}$ and $\Delta R_{ij}$ denote the invariant mass and the angular distance between two reconstructed objects $i$ and $j$, $p_{Ti}$ denotes the transverse momentum of an object $i$, $\met$ denotes the missing transverse energy, and $S_T$ is the scalar sum of transverse momenta of the reconstructed jets and the two same-sign leptons. 

\subsection{Machine learning aspects}\label{sec:MLaspects}

\begin{figure*}[t]
	\centering
	\begin{center}
		\includegraphics[width=0.89\textwidth,clip]{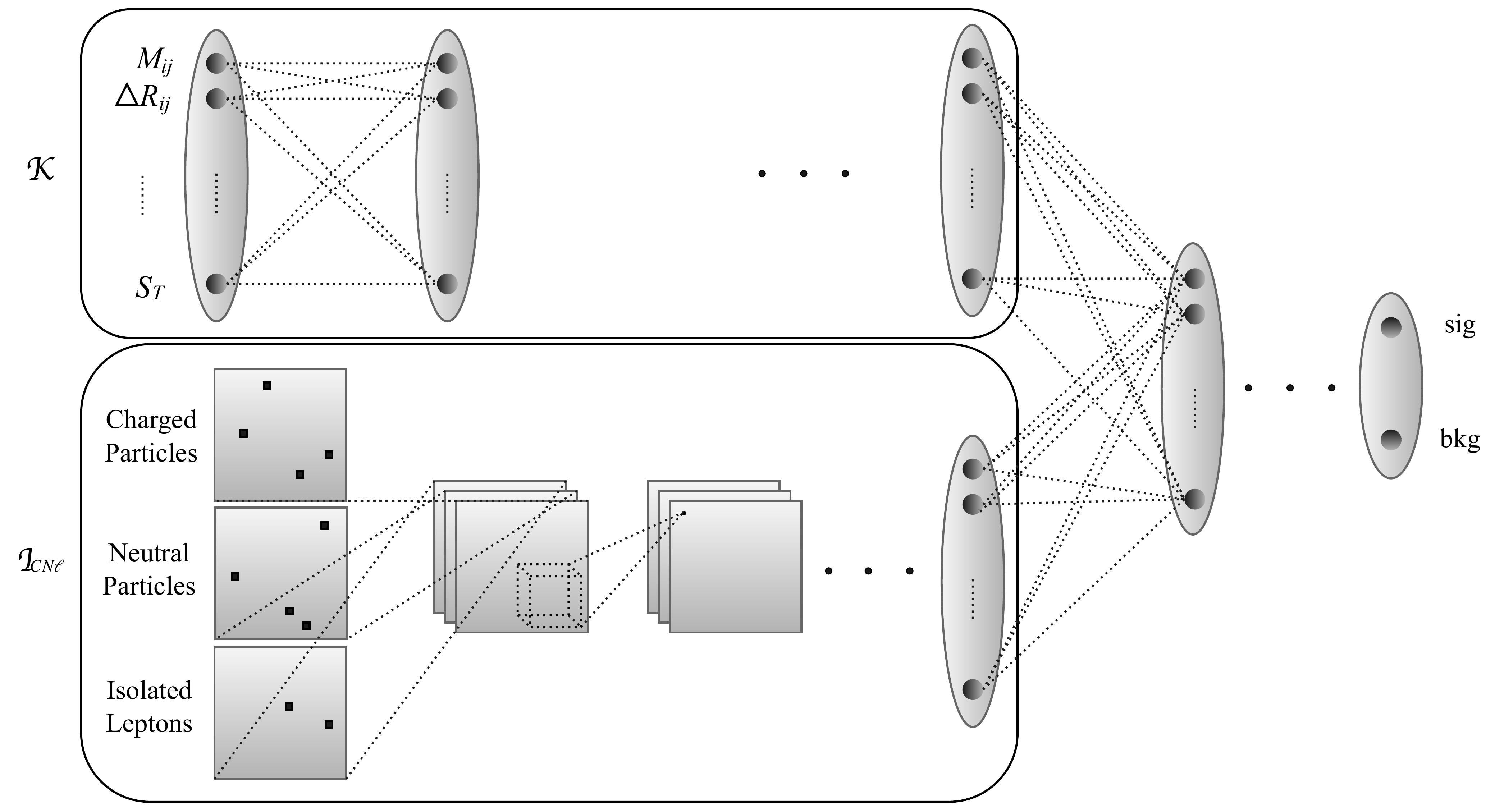}
		\caption{A schematic architecture of the neural networks used in this paper. The separate DNN chain in the upper panel is used only when the kinematic variables are included.}
		\label{fig:NN_CNLK} 
	\end{center}
\end{figure*}

Using the jet images and kinematic data defined in the previous sections, we train neural networks to differentiate the signal from its SM backgrounds. We have implemented a number of different networks which are described in detail and compared in \cref{sec:appen}.
For some of the networks we only use part of the available data to assess which data sets yield the highest discriminatory power:
\begin{itemize}
	\item $\mathcal K$: Fully connected layer using only the kinematic data. 
	\item $\mathcal I_{C\ell}/\mathcal I_{CN\ell}$: Convolutional neural network~\cite{LeCun:2012} using charged (and neutral) hadron images and lepton images.
	\item $\mathcal I_{C(N)\ell}+\mathcal K$: CNN using charged (and neutral) hadron images and lepton images, combined with a FC using the kinematic data. The flattened part of the CNN and the kinematic FC are interfaced to produce a combined output. 
\end{itemize}
The schematic structure of the best-performing network $\mathcal I_{CN\ell}+\mathcal K$ is illustrated in \cref{fig:NN_CNLK}. 
The networks are implemented in the \texttt{PyTorch}~\cite{paszke2019pytorch} framework.
Further details on the network architectures are provided in \cref{sec:appen}.

The data set consists of 130,000 events, which are split into training (64\%), validation (16\%), and test (20\%) sets. 
When training the network, the entire training data is split into a smaller number of samples (referred to as mini-batches) so that, in each iteration, the network is trained with a randomly chosen mini-batch. 
It is iteratively trained with the Adam optimiser~\cite{Kingma:2014vow} to update the hyper-parameters (referred to as weights) of the network in a way to minimise the cross entropy loss function.

The validation set is used to monitor the generalisation error and choose the best model parameters as those with the lowest validation loss, and the test set is exclusively used for final evaluation.
All three data sets are comprised to equal parts of signal and background events.
The number of events of the different backgrounds is proportional to the respective cross sections after the preselection cuts.
Unless stated otherwise, we train and validate the network with a signal data set that contains several mass hypotheses: $m_S=300$~GeV to 800~GeV in steps of 50~GeV, each appearing with equal weight.
For the test sets, only a single mass $m_S$ is used, $m_S=300$~GeV to 1000~GeV in steps of 50~GeV.

While a full treatment of the systematic uncertainties is beyond the scope of this article, we have control over the systematics that are due to the machine learning: The limited number of Monte Carlo events makes the results susceptible to statistical fluctuations. Another contribution is the stochastic nature of the initialisation of the network parameters and the gradient descent. 
In order to take these uncertainties into account, we perform 20 independent training runs for each network and evaluate them separately. 
In the following, the uncertainty bands indicate the $1\sigma$ fluctuations between the different runs.

\section{Results}
\label{sec:results}

\begin{figure}[t]
	\centering
	\begin{subfigure}[position]{0.31\textwidth}
		\includegraphics[width=\linewidth]{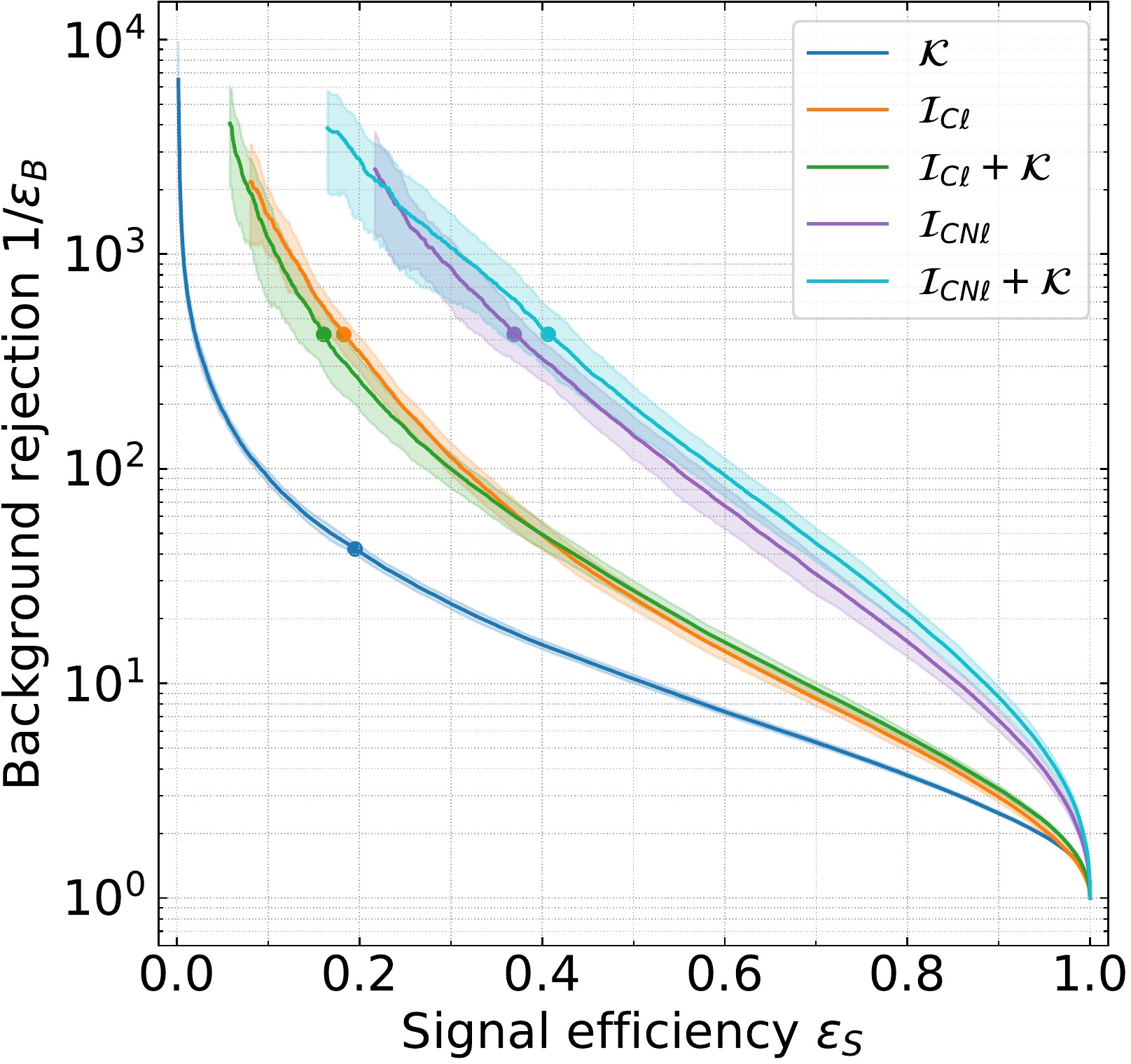}
		\caption{$m_S=300$~GeV}
	\end{subfigure}\quad
	\begin{subfigure}[position]{0.31\textwidth}
		\includegraphics[width=\linewidth]{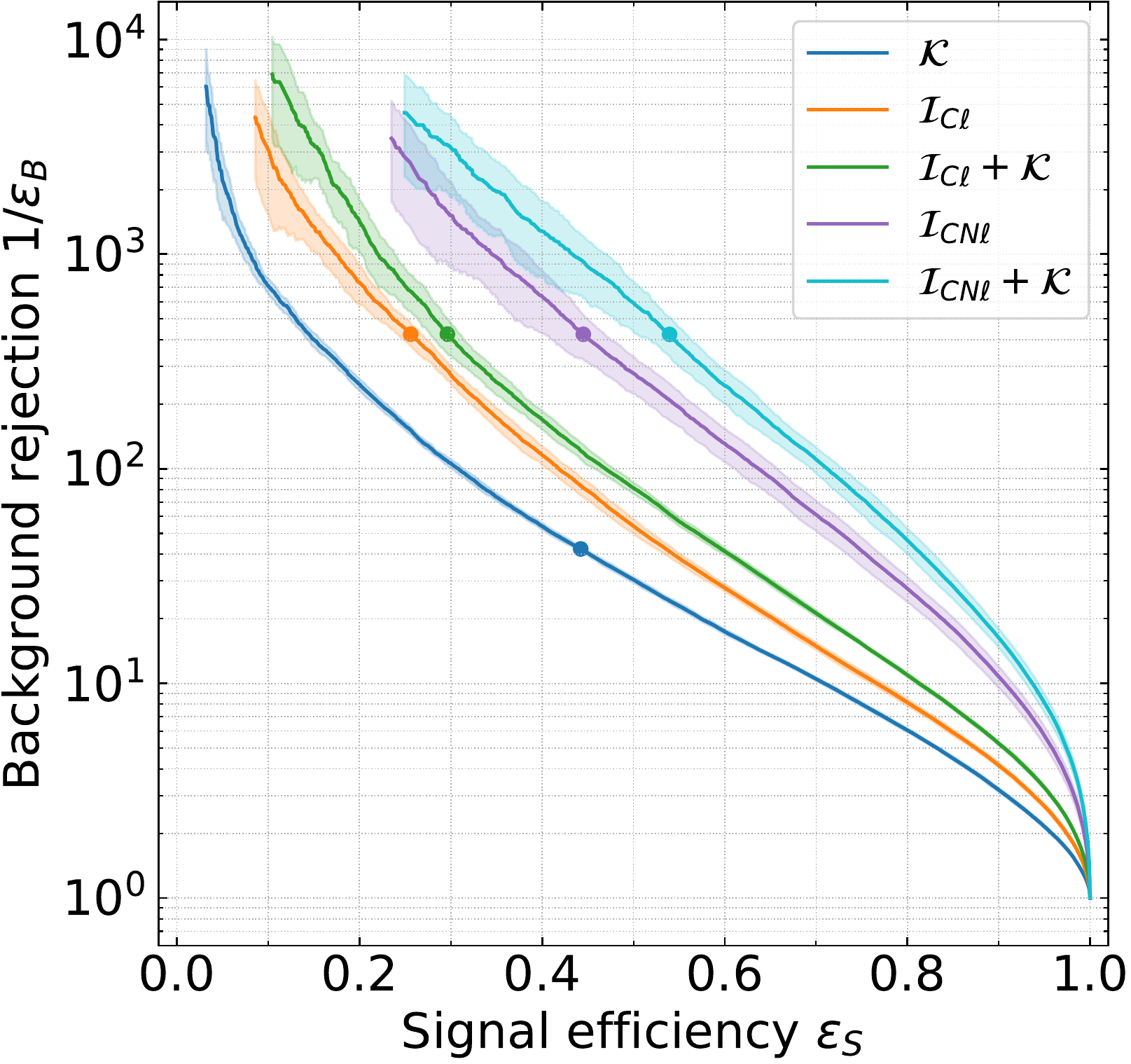}
		\caption{$m_S=550$~GeV}
	\end{subfigure}\quad
	\begin{subfigure}[position]{0.31\textwidth}
		\includegraphics[width=\linewidth]{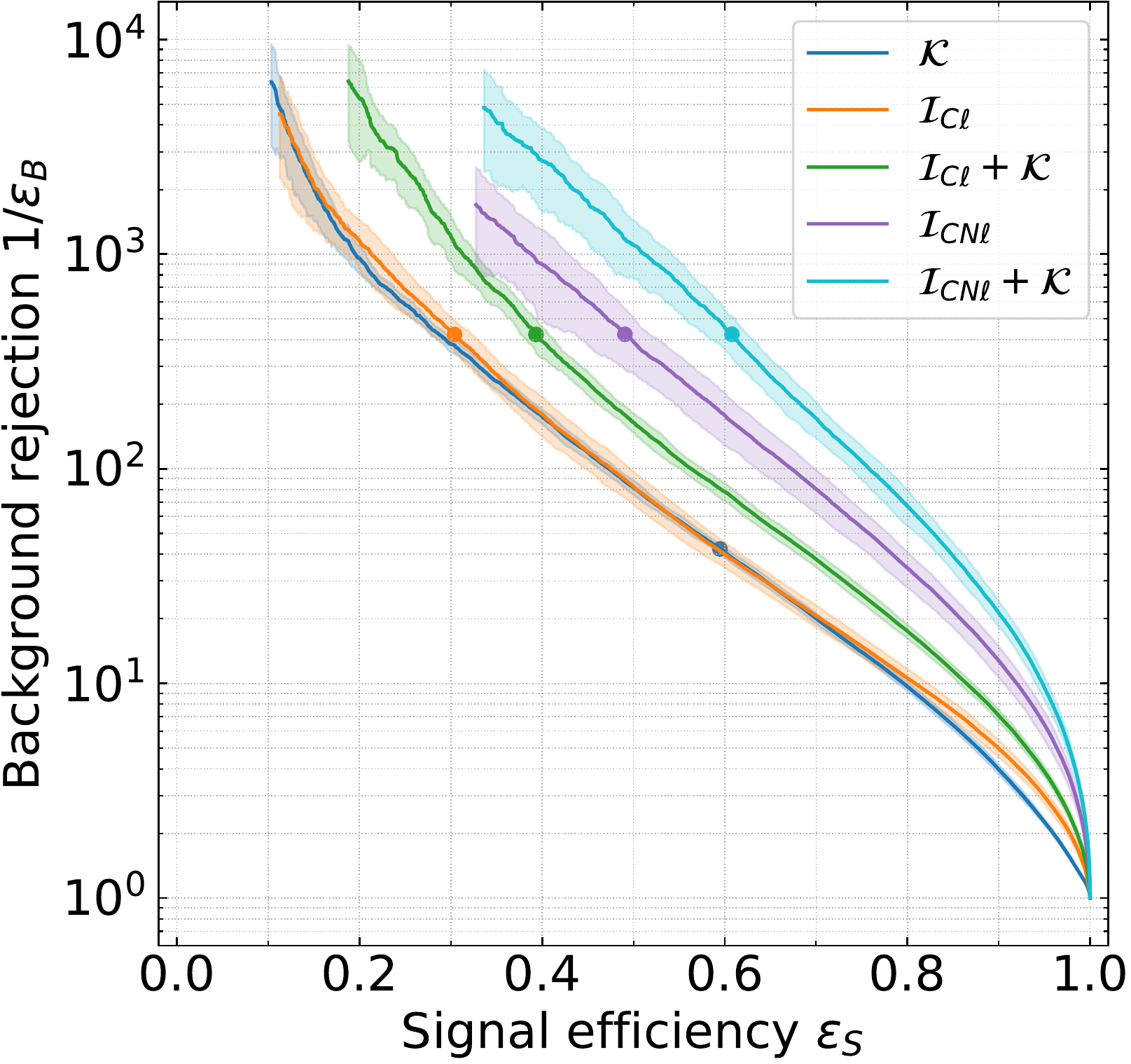}
		\caption{$m_S=800$~GeV}
	\end{subfigure}
	\caption{Comparison of network performances with ROC curves. The markers indicate the working points used in the following analysis.}
	\label{fig:roc}
\end{figure}

We start by comparing the performance of the trained networks as classifiers.
To this end, the receiver operating characteristic  (ROC) curves in \cref{fig:roc} show how many background events can be rejected for a given signal efficiency.
As expected, the performance increases from $\mathcal K$ to $\mathcal I_{CN\ell} + \mathcal K$ as more data are used.
All networks perform better at higher signal masses $m_S$, but this effect is largest for the kinematic data: $\mathcal K$ has the largest boost in performance, and only for higher $m_S$ does the inclusion of kinematic data significantly improve the image-based classification.
This can be explained by the increased $p_T$ of the final state particles.

When calculating the $2\sigma$ exclusion bound~\cite{Cowan:2010js}, we require:
\begin{equation}
Z_\mathrm{exc} \equiv
\sqrt{-2\,\ln\bigg(\frac{L(S\!+\!B | B )}{L( B | B)}\bigg)} \geq 1.64 ,
\;\;\;\;\; \text{with}\;\;\;
L(x |n) =  \frac{x^{n}}{n !} e^{-x}\,,
\label{Eq:SigExc} 
\end{equation}
where $L(x |n)$ is the likelihood of observing $n$ events when $x$ events were expected, and $S$ and $B$ are the number of signal and background events, respectively.
For achieving a $5\sigma$ expected discovery reach, we require
\begin{equation}
Z_\mathrm{dis} \equiv
\sqrt{-2\,\ln\bigg(\frac{L(B | S\!+\!B)}{L( S\!+\!B| S\!+\!B)}\bigg)} \geq 5 \,.
\label{Eq:SigDis} 
\end{equation}

When putting a cut on the neural network score (NN score), it is common practice to choose the cut that maximizes \cref{Eq:SigDis} \cite{Huang:2022rne}. However, our networks are such strong classifiers, that this method regularly leads to cuts with few or even less than 1 background event remaining, for which a statistical treatment becomes challenging. We therefore conservatively choose our NN score cut by demanding a fixed amount of background events.
Details which justify this approach are provided in \cref{sec:appcut}.
Once the NN score cut is fixed, we determine the $5\sigma/2\sigma$ cross sections by rescaling the cross section iteratively  until \cref{Eq:SigDis,Eq:SigExc} are satisfied.

\begin{figure}[t]
	\centering
	\begin{subfigure}{0.47\linewidth}
		\includegraphics[width=\linewidth]{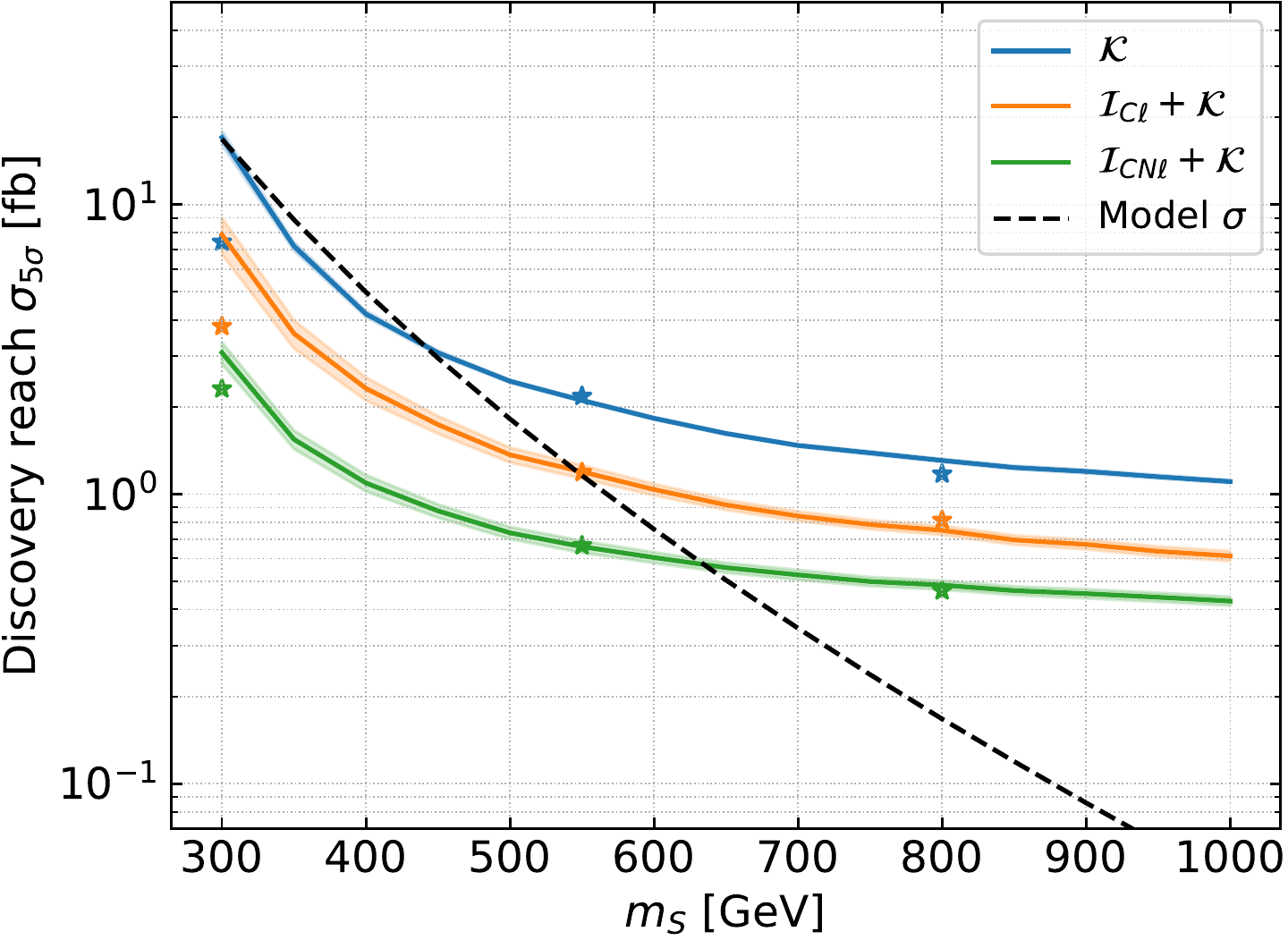}
		\caption{Discovery reach of $S^{++} S^{--}$}\label{fig:xsmain1}
	\end{subfigure}\quad
	\begin{subfigure}{0.47\linewidth}
		\includegraphics[width=\linewidth]{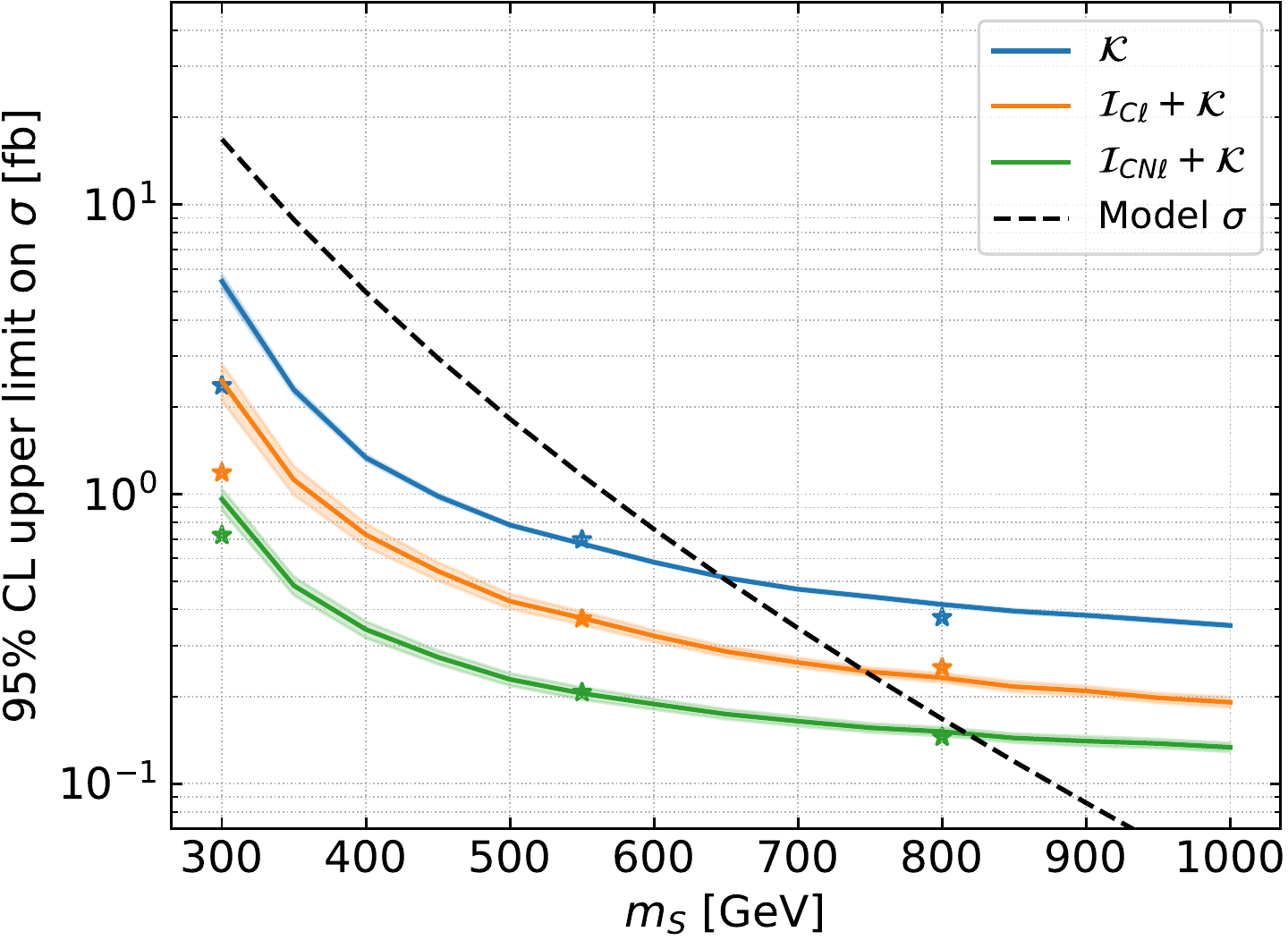}
		\caption{Expected upper limit of $S^{++} S^{--}$}\label{fig:xsmain2}
	\end{subfigure} \vspace{2ex}
	
	\begin{subfigure}{0.47\linewidth}
		\includegraphics[width=\linewidth]{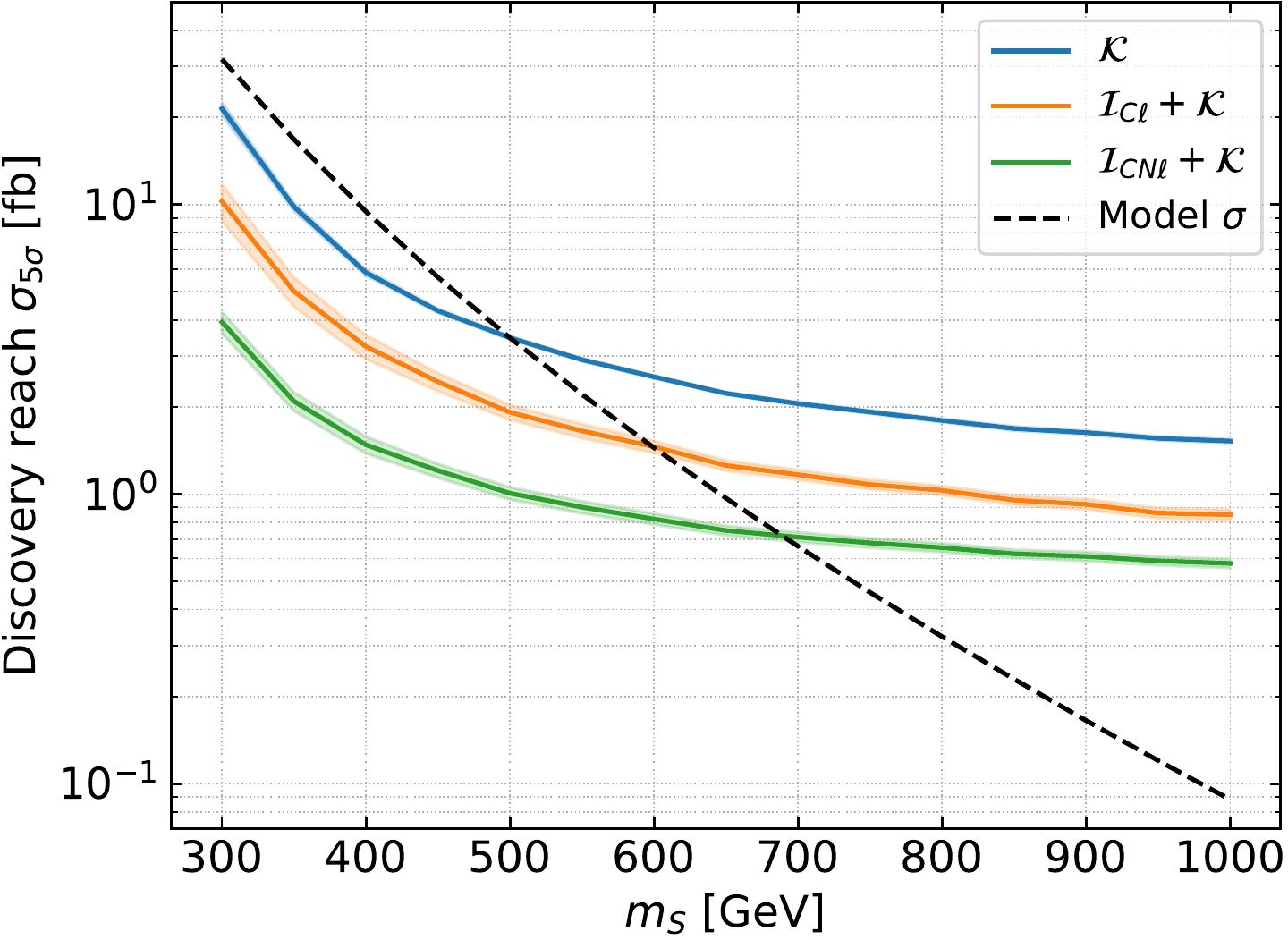}
		\caption{Discovery reach of $S^{++} S^{--}$ and $S^{\pm\pm} S^\mp$}\label{fig:xsmain3}
	\end{subfigure}\quad
	\begin{subfigure}{0.47\linewidth}
		\includegraphics[width=\linewidth]{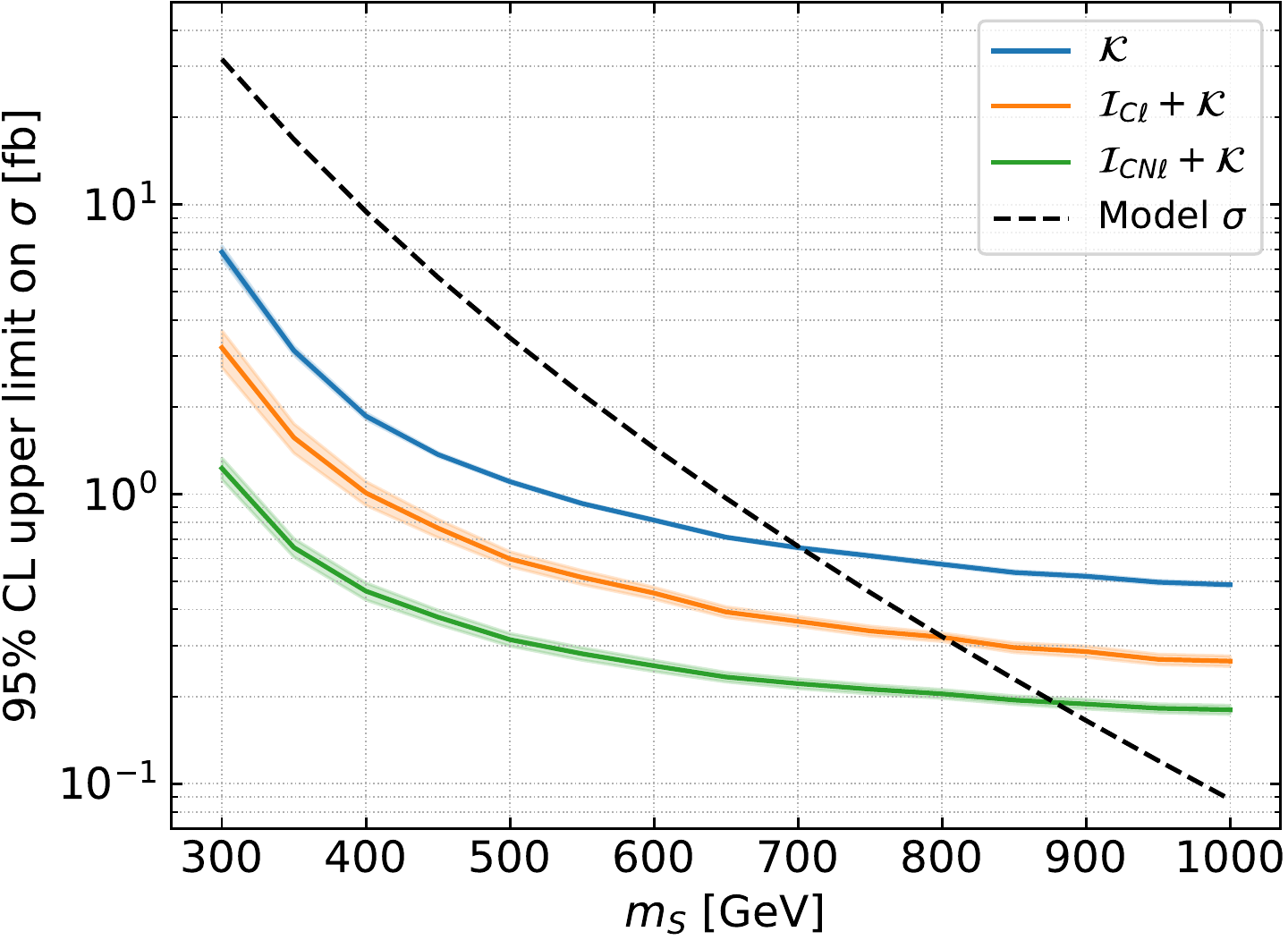}
		\caption{Expected upper limit of $S^{++} S^{--}$ \& $S^{\pm\pm} S^\mp$}\label{fig:xsmain4}
	\end{subfigure}
	\caption{Expected discovery reach (left) and exclusion limit (right) of $S^{++}$ production at the HL-LHC (3~ab$^{-1}$) for different network architectures. In the first row simulated events of only $S^{++} S^{--}$ pair production are taken into account. In the second row simulated events of $S^{++} S^{--}$ and $S^{\pm\pm} S^\mp$ production with $m_{S^{++}}= m_{S^+}$ and the corresponding cross sections from \cref{fig:Crosssection} are taken into account. The dashed lines indicate the reference cross sections in the SU(5)/SO(5) model of $S^{++} S^{--}$ (top row) and the sum of the cross sections of $S^{++} S^{--}$ and $S^{\pm\pm} S^\mp$ (bottom row) at 14~TeV. For the solid lines, the networks were trained on multiple masses simultaneously. The points marked by stars were trained using only the respective masses.}\label{fig:xsmain}
\end{figure}

The expected discovery reach and upper limits for the networks $\mathcal K$, $\mathcal I_{C\ell}+\mathcal K$, and $\mathcal I_{CN\ell}+\mathcal K$ are shown in \cref{fig:xsmain1,fig:xsmain2}, where the NN score cuts were placed to obtain 50, 5, and 5 background events after the cut, respectively, and we assume an integrated luminosity of 3~ab$^{-1}$. 
The corresponding signal efficiencies can be read off from the markers in \cref{fig:roc}. 
The black dashed line indicates the 14~TeV pair production cross section in the SU(5)/SO(5) model.
Using only kinematic data, the signal process can just barely be discovered up to $440$~GeV and excluded up to $650$~GeV. If we combine them with charged jet and lepton images, however, the signal can be discovered up to $550$~GeV and excluded to $750$~GeV.
If advances in pileup suppression also allow the usage of neutral jet images, the discovery reach can be increased by another $90$~GeV while the projected exclusion is increased by $70$~GeV.

Discovery reach and expected bounds on $\sigma$ become significantly weaker for low masses. 
This can be partially remedied by choosing a different training data set. The solid lines were obtained by training on a combined data set containing equal amounts of signal events with masses between 300 and 800~GeV in steps of 50~GeV. 
However, the stars in \cref{fig:xsmain1,fig:xsmain2}  show the bound that can be reached if the networks are trained only on signal events with the mass they are evaluated at. 
This has little impact for $\mathcal I_{CN\ell}+\mathcal K$, but noticeably improves the bounds for $\mathcal K$ and to a lesser extent also $\mathcal I_{C\ell}+\mathcal K$ at $m_S=300$~GeV.
At higher masses, training on a single mass is less beneficial or even slightly detrimental compared to the combined data set.
We explore the impact of the training data set further in \cref{sec:apptrainingset}.

In the second row of \cref{fig:xsmain}, also the singly charged scalar is taken into account. The solid lines are now discovery reach (left) and exclusion limit (right) for the combined $S^{++} S^{--}$ and $S^{\pm\pm} S^\mp$ production with the relative cross sections shown in \cref{fig:Crosssection}. The dashed line is the sum of the two cross sections, so that an intersection indicates discovery/exclusion of both processes.
By including the second process, both the discovery reach and the exclusion limit are increased by about 50~GeV for all three networks.

\begin{figure}[t]
	\centering
	\includegraphics[width=0.47\textwidth]{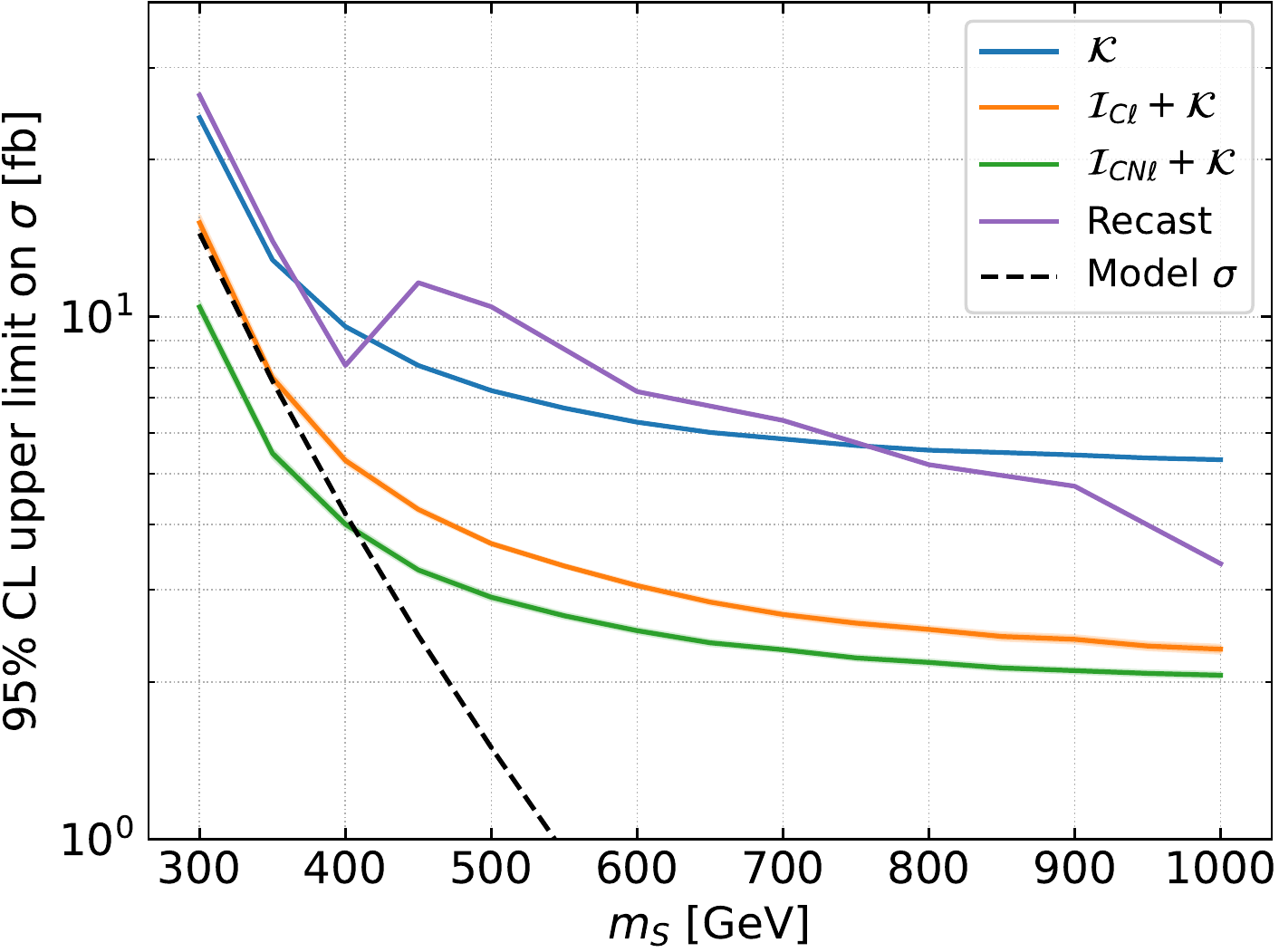}
	\caption{Expected exclusion limit of $S^{++} S^{--}$ pair production at the LHC with $\mathcal L_\mathrm{int}=139$~fb$^{-1}$ for different network architectures. The recast bounds are taken from Ref.~\cite{Cacciapaglia:2022bax}. The black line indicates the 13~TeV reference cross sections in the SU(5)/SO(5) model.}\label{fig:xs139}
\end{figure}

Having now seen the bounds from the neural network approach, it is natural to ask how this compares to conventional cut-and-count methods.
No direct search for the present final state has been performed yet, but in Ref.~\cite{Cacciapaglia:2022bax} an upper limit on the cross section is derived from the recast of an ATLAS search for $R$-parity violating supersymmetry using the full \mbox{Run-2} data set~\cite{ATLAS:2021fbt}. 
The recast results are compared with the bounds derived from the neural networks with 139~fb$^{-1}$ in \cref{fig:xs139}.
Naturally, a dedicated search would provide stronger bounds than the recast, so we cannot draw quantitative conclusions from \cref{fig:xs139}. 
However, it is evident that the performance of the network based on purely kinematic information is comparable to the recast of \cite{ATLAS:2021fbt} for low $m_S$. The signal regions which yield the dominant recast bound demand same-sign leptons and multiple $b$-tagged jets and are thus similar to our preselection criteria, and the selection criteria of \cite{ATLAS:2021fbt} are based purely on kinematic information. Inclusion of jet image data of charged and potentially neutral jet images processed by a CNN yield a substantial improvement in discrimination power in our study.

\section{Conclusion and outlook}
\label{sec:conclusion}

The search for new physics is and remains one of the main aims of the LHC. Composite Higgs models with extended scalar sectors are theoretically well motivated, their signatures are far from being ruled out, and the LHC can explore untested parameter space at its high luminosity run if dedicated searches are performed. In this article, we proposed a search strategy for a doubly charged scalar $S^{++}$ which is pair produced and which decays into $W^+t\bar{b}$. An  $S^{++}$ with this decay mode is predicted -- for example -- by composite Higgs models with an underlying fermionic description based on the global symmetry breaking pattern SU(5)$\rightarrow$SO(5).

In our proposed search, we focus on the process $p p \rightarrow S^{++}S^{--}\rightarrow W^+t\bar{b}W^-\bar{t}b$. This process is not constrained by current LHC searches \cite{Cacciapaglia:2022bax} and our proposal is aiming to close this gap. We target the same-sign lepton final state of this channel.  We implement deep neural networks which aim to optimise the discrimination power between signal and SM backgrounds based on ($\mathcal{K}$) kinematic information of reconstructed jets and leptons; ($\mathcal{K}+\mathcal{I}_{C\ell}$) additional information from jet images of charged jets and leptons; ($\mathcal{K}+\mathcal{I}_{CN\ell}$) yet additional information from jet images of neutral jets,  as defined in detail in \cref{sec:input}. As a main result (shown in \cref{fig:xsmain}) we obtain projected discovery and exclusion bounds at the high luminosity run of the LHC for (A) doubly charged scalars in the process $p p \rightarrow S^{++}S^{--}\rightarrow W^+t\bar{b}W^-\bar{t}b$ and (B) models with doubly and singly charged scalars, where the latter is contributing to the signal region via the process $p p \rightarrow S^{++}S^{-} \rightarrow W^+t\bar{b}\bar{t}b$ and its conjugate process. We find that the use of convolutional neural networks on charged jet and lepton images substantially increases the discovery and exclusion potential, with projected discovery reach of $m_S \leq 550$~GeV and $m_S \leq 600$~GeV for the scenarios (A) and (B) and a projected exclusion potential up to $m_S \leq 750$~GeV and $m_S \leq 800$~GeV, respectively. A further improvement can be achieved when including neutral jet images into the data set, which can raise projected discovery reaches to $m_S \leq 640$~GeV and $m_S \leq 690$~GeV and projected exclusion potential to $m_S \leq 820$~GeV and $m_S \leq 880$~GeV. However, this latter improvement relies on sufficiently controlling pileup effects. We do not study pileup effects in this article and instead present results for $\mathcal{K}+\mathcal{I}_{C\ell}$ data as conservative and results for $\mathcal{K}+\mathcal{I}_{CN\ell}$ data as a potentially realisable but optimistic scenario.

For this study, we implemented a variety of neural networks which are discussed and compared in \cref{sec:appen} to which we refer especially readers interested in the machine learning aspects of this proposed BSM search. \cref{sec:appeval} contains additional details related to the choice of the cut on the NN score and the composition of the training data set. We consider the information therein methodologically relevant not only for the study presented here, but more generally for BSM searches with unknown mass of the BSM particle and with potentially very low SM backgrounds after NN score cuts.

This study demonstrates that the use of machine learning techniques in general, and the use of jet images and CNNs in particular, provides a large potential to improve the sensitivity of LHC searches for BSM final states with a lot of hadronic activity. The channel $p p \rightarrow S^{++}S^{--}\rightarrow W^+t\bar{b}W^-\bar{t}b$ in the same-sign lepton final state provides one example. We expect that similar improvements are achievable for other final states like Drell-Yan pair produced BSM scalars, pair produced vector-like quarks with exotic decays, or signals from R-parity violating supersymmetry to just name a few examples which deserve further investigation.

\section*{Acknowledgements}

We thank Raimund Str\"ohmer for discussions. 
This work has been supported by the international cooperation program ``GEnKO'' managed by the National Research Foundation of Korea (No. 2022K2A9A2A15000153, FY2022) and DAAD, P33 - projekt-id 57608518. This work is supported by the Center for Advanced Computation at Korea Institute for Advanced Study. 
T.F.\ is supported by a KIAS Individual Grant (AP083701) via the Center for AI and Natural Sciences at the Korea Institute for Advanced Study. P.K.\ is supported in part by a  KIAS Individual Grant No. PG021403, and by National Research Foundation of Korea (NRF) Research Grant NRF- 2019R1A2C3005009. M.K.\ is supported by the ``Studienstiftung des deutschen Volkes''.
J.K.\ is supported by the National Research Foundation of Korea (NRF) grant funded by the Korea government (MSIT) (No. 2021R1C1C1005076).
J.K.\ acknowledges the hospitality at APCTP where part of this work was done.

\appendix

\section{Neural network architectures}
\label{sec:appen}
Machine learning can bring significant improvements to results, but it is important to understand how to effectively utilise them. For this work, we tested various types of data representations and neural network architectures and found that a combination of kinematic variables together with low level calorimeter information represented in terms of jet images  yields the best results. In this section, we describe several neural network architectures we have attempted and evaluate the performance of each network by analysing the ROC curves. This analysis uses only a data set with the mass of the signal $m_S=400$~GeV, and we define two additional kinematic variables.
\begin{eqnarray} 
&\mathcal K_{p_{\mu}}^{\rm (vis)} = \{ p_\mu (\ell_1), p_\mu (\ell_2), p_\mu (b_1), p_\mu (b_2), p_\mu (b_3), p_\mu (j_1), p_\mu (j_2), p_\mu (j_3) \} \, ,
\label{eq:vis_4momenta} \\
&\mathcal K_{2} = \bigcup_{i \neq j} M_{ij} \cup \bigcup_{i \neq j }\Delta R_{ij}\;  ,\label{eq:2kin}
\end{eqnarray}
where $\ell_1$, $\ell_2$, $b_i$, and $j_i$ denote two same-sign leptons, $i$-th leading $b$-tagged jets, and $i$-th leading non-$b$-tagged jets respectively.

\subsection{Fully connected deep neural networks}
\label{sec:FC}
\begin{figure*}[t]
	\centering
	\begin{center}
		\includegraphics[width=0.5\textwidth,clip]{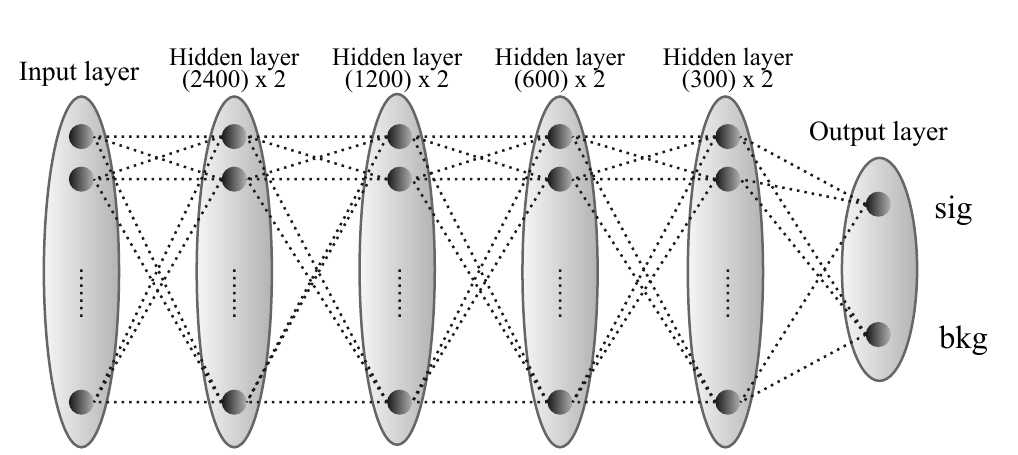}
		\caption{A schematic architecture of the fully-connected NN (FC) used in this paper.}
		\label{fig:FC} 
	\end{center}
\end{figure*}

Fully connected layer (FC) neural networks are  simple structures that can easily be used when an input layer is composed of the four-momenta of reconstructed particles or combinations of kinematic variables. As shown in \cref{fig:FC}, FCs are characterised by all neurons in each layer being connected to all neurons in the next layer. In the FC we implemented, there are 8 hidden layers, with the number of neurons decreasing from 2400 to 300, and a last hidden layer is connected to an output layer. The output layer consists of two neurons representing  background or signal, respectively. The values associated with these two neurons are denoted as $\hat{p}_k$, where $k$ is 0 and 1 for a background and a signal, respectively. 
We normalise the values $\hat{p}_k$ as $p_{k} = e^{\hat{p}_{k}}/\sum e^{\hat{p}_{k}}$ using a softmax function, where $p_0$ represents the ``probability to be a background'' that the network assigns to this event, and analogous with $p_1$ and signal.
We use the  Rectified Linear Unit (ReLU) \cite{pmlr-v15-glorot11a} activation function between hidden layers. When training the network, we use a mini-batch size of 20 and a learning rate of $5 \times 10^{-7}$ of the Adam optimiser to minimise the cross entropy loss function
\begin{eqnarray}
L = -y_k \log p_k - (1-y_k) \log (1-p_k) \;,
\end{eqnarray} 
where a background and a signal event are labeled as $y_0 = 0$ and $y_1 = 1$, respectively. We include the $L_2$ regularisation\footnote{
	The $L_2$ regularisation shifts the loss function by $L \rightarrow L +  \frac{1}{2} \lambda \|\textbf{W}\|^2$ where $\textbf{W}$ represents all weights, and $\lambda$ denotes {\tt weight\_decay}. In this paper, we have applied the $L_2$ regularisation to all neural networks.} terms to the loss function with {\tt weight\_decay}=$2 \times 10^{-9}$ to alleviate the problem of data overfitting.
The learning rate and \texttt{weight\_decay} are the same for the cases of $\mathcal K_{p_{\mu}}^{\rm vis}$, $K_{p_{\mu}}^{\rm vis} + \mathcal K_2$, and $\mathcal K$. In addition, we use a validation set to make sure that the neural network is not overfitted. We train the model for 100 epochs. If the validation loss does not reach a new minimum within 50 epochs, the training is stopped and the parameters of the epoch with the minimal validation loss are saved\footnote{
	Unless otherwise stated, the activation function, the type of optimiser, the loss function, the configuration of an output layer, and the use of a validation set are the same for all other neural networks.
}.

\subsection{Convolutional neural networks}
\label{sec:CNN}
\begin{figure*}[t]
	\centering
	\begin{center}
		\includegraphics[width=1.\textwidth,clip]{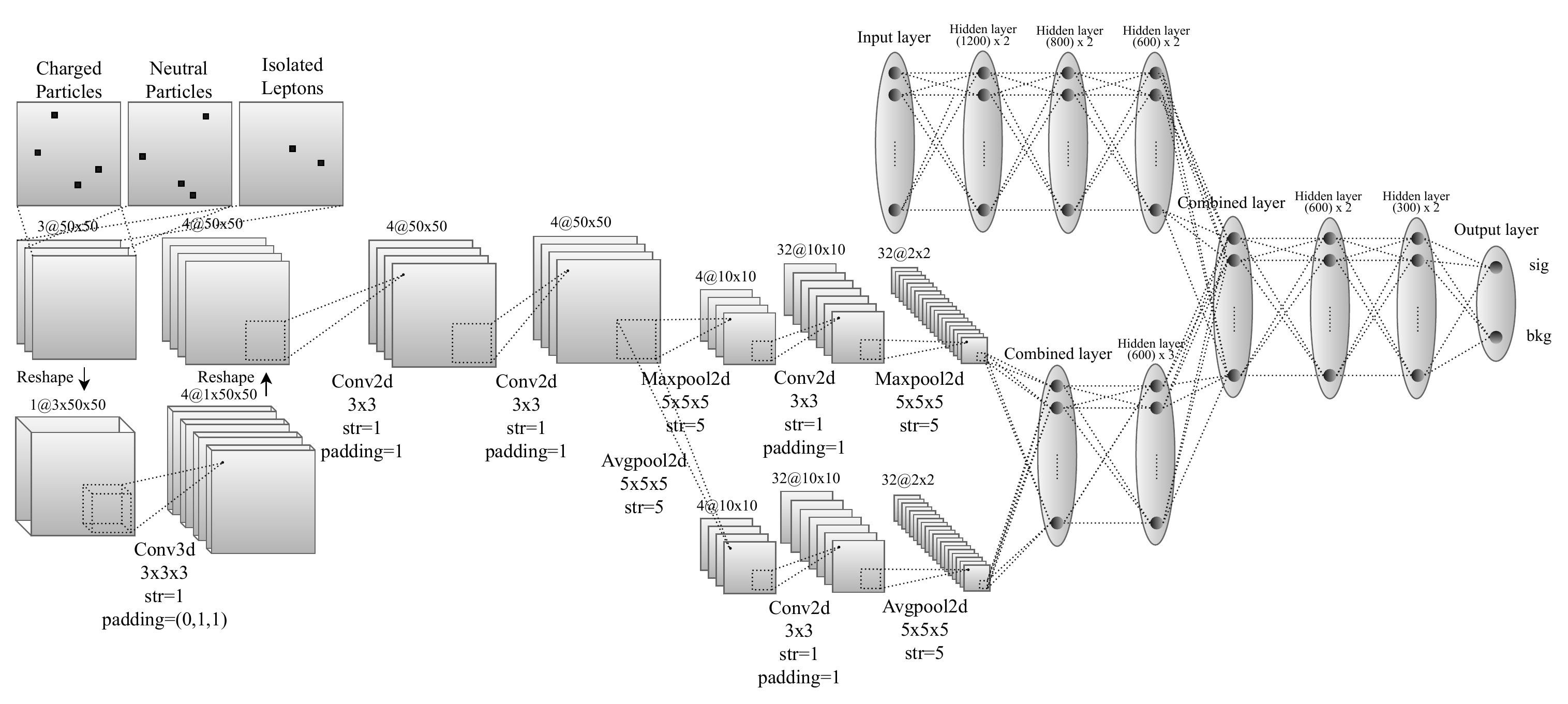}
		\caption{A schematic CNN architecture  used in this article. The separate FC chain in the right-upper corner is used only when kinematic variables are included.}
		\label{fig:CNN} 
	\end{center}
\end{figure*}

CNNs are a class of machine learning algorithms commonly used for image recognition tasks, and they are useful when a final state is represented as a set of images.
An input is a 3D image of $\mathcal I_{CN\ell}$ ($\mathcal I_{C\ell}$) 
whose dimension is given by $ 3 \times 50 \times 50  $ ($ 2 \times 50 \times 50 $) where 3 (2) denotes the number of image layers in \cref{eq:3images} (\cref{eq:2images}).

To fully exploit spatial correlations among different layers, it is necessary to utilise a 3D convolution. 
For this operation, we need to add a singleton dimension to reshape the image dimension as $ 1 \times 3 \times 50 \times 50  $ ($ 1 \times 2 \times 50 \times 50  $), where each entry is identified as a channel, a depth, a height, and a width respectively.
Then, we apply the 3D convolution with kernel size of $3 \times 3 \times 3$ ($2 \times 3 \times 3$), stride of 1, padding size of $(0,1,1)$, and 4 feature maps. After the 3D convolution, we apply the batch normalisation and the ReLU activation function, and the resulting image dimension becomes $ 4 \times 1 \times 50 \times 50 $. To remove the singleton ($i.e.$ redundant) dimension in the second entry, we reshape the image size again to make it as $ 4 \times 50 \times 50$.

Next, we apply a 2D convolution twice using the kernel size of $3 \times 3$, the stride of 1, the padding size of 1, and 4 feature maps. Throughout this paper, for all 2D convolutions, padding size is fixed to 1, and after each 2D convolution we apply the batch normalisation and the ReLU activation function.

The pooling method used in CNNs takes the most representative pixel in each region of a feature map. For example, a max (average) pooling operation keeps the maximum (average) value within a feature map, while discarding the other remaining information. This is how most CNNs reduce a dimension of an input image data through a series of combined convolutional and pooling operations to learn important characteristics of a given image in various scales.
As shown in \cref{fig:CNN}, we design a two-pronged architecture to include both the max and the average pooling operations with a kernel size of $5 \times 5$ and a stride of 5, which reduces the image dimension down to $4 \times 10 \times 10$. Each branch is followed by the 2D convolution and the pooling again, until its size is reduced to $32 \times 2 \times 2$. We flatten out the final image into a one-dimensional strip to combine it with 3 hidden layers with 600 neurons each. 

A neural network that combines kinematic variables with a CNN architecture has been demonstrated in Ref.~\cite{Huang:2022rne}. In a similar way, we construct a separate FC with 6 hidden layers, with the number of neurons decreasing from 1200 to 600 (as shown in the upper-right corner of the \cref{fig:CNN}). The last FC layer with 600 neurons is combined with the last hidden layer of the CNN. The combined layer is followed by 4 hidden layers with the number of neurons decreasing from 600 to 300. The last hidden layer is then connected to the output layer. The network is trained using a mini-batch size of 20. The learning rate is set to $5 \times 10^{-4}$ for $\mathcal I_{C\ell}$ and $\mathcal I_{CN\ell}+\mathcal K$, and $3 \times 10^{-4}$ for $\mathcal I_{CN\ell}$ and $\mathcal I_{C\ell}+\mathcal K$. We use \texttt{weight\_decay}=$2 \times 10^{-4}$ for $\mathcal I_{C\ell}+\mathcal K$ and $3 \times 10^{-4}$ for the remaining networks.

\subsection{Capsule neural networks}
\label{sec:CapsNet}
\begin{figure*}[t]
	\centering
	\begin{center}
		\includegraphics[width=1.\textwidth,clip]{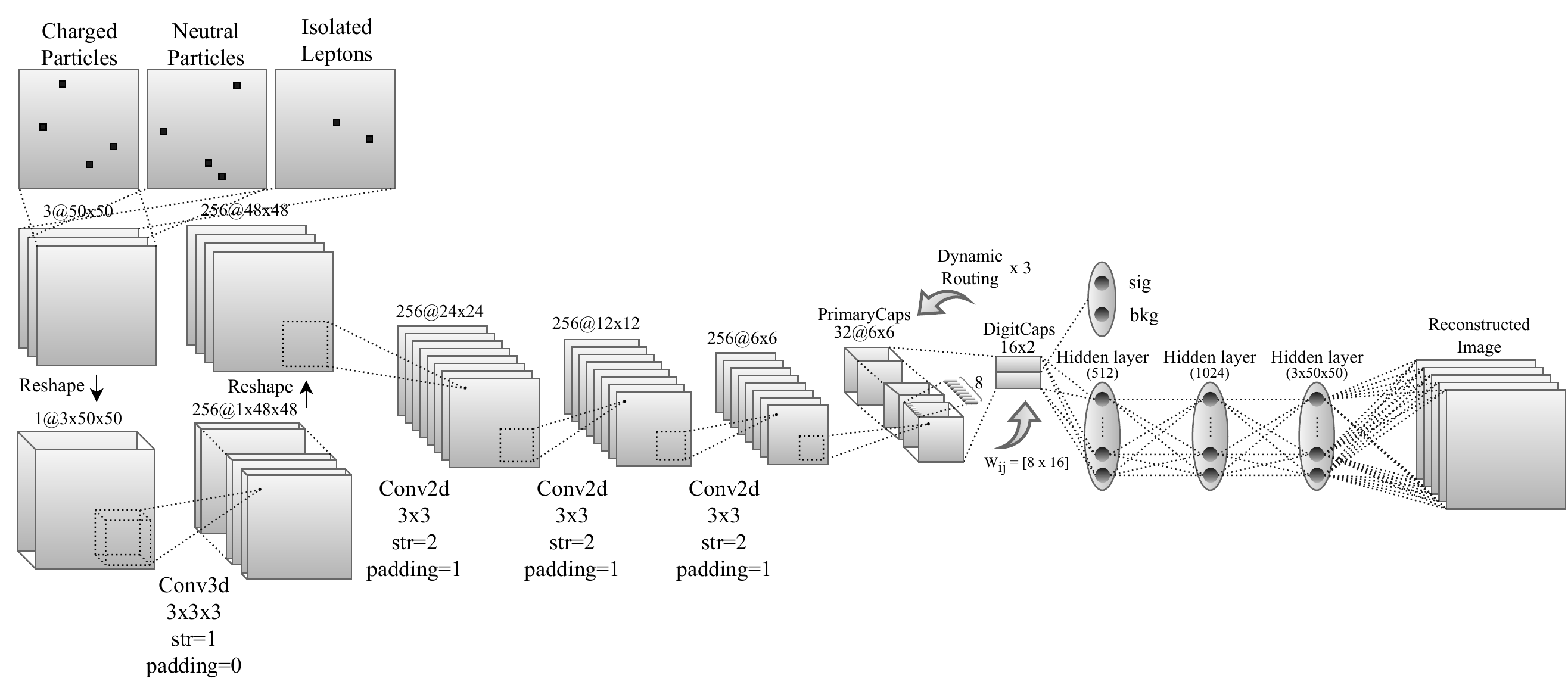}
		\caption{A schematic architecture of the CapsNet used in this paper.}
		\label{fig:CapsNet} 
	\end{center}
\end{figure*}

As explained in the previous subsection, the pooling operation in CNNs is the mechanism to reduce the image dimension to learn important features in various scales.
However, a major problem associated with this pooling operation is that it loses information by throwing away other pixels. Eventually, it accompanies the loss of spatial information about where things are. Therefore, CNNs are agnostic to the geometric correlations among the pixels at higher and lower levels. A capsule neural network (CapsNet)~\cite{NIPS2017_2cad8fa4,Diefenbacher:2019ezd} was proposed to address this problem by introducing a concept of dynamic routing, which compensates for the shortcomings of the pooling operations. 
For dedicated readers, we refer to Ref.~\cite{Huang:2022rne} for more detailed descriptions of the CapsNet and its variants. 
Although it was originally proposed to improve CNNs, the previous study~\cite{Huang:2022rne} demonstrated that the performance of the CapsNet was worse than CNNs in general.
We suppose that it would have been attributed to sparse jet images in their example of the double Higgs production  $(h\to b{\bar b})(h\to W^\pm W^{*\mp}\to \ell^+\nu_\ell {\ell^\prime}^- \bar{\nu}_{\ell^\prime})$.
On the other hand, in our signal process, the busier final state delivers denser jet images with more activated pixels.
Ample features would appear in jet images that the CapsNet might take advantage of.
This motivates us to apply the CapsNet to our image data as a comparison with CNNs to see how much we gain from using a different type of image-based neural network.

Given a 3D image data set $\mathcal I_{CN\ell}$ ($\mathcal I_{C\ell}$), we first apply a 3D convolution with a kernel size of $ 3 \times 3 \times 3  $ ($ 2 \times 3 \times 3 $), a stride of 1, a padding of 0, and out channels of 256. Then, we apply a 2D convolution with a kernel size of $ 3 \times 3 $, a stride of 2, and a padding of 1 until the image dimension is reduced down to $ 256 \times 6 \times 6 $. The output neurons of the 2D convolution are reshaped to form $6 \times 6 \times 32 = 1152$ primary capsule vectors of length 8, which contain lower-level information of the input image. These primary capsules are connected to two 16-dimensional higher-level capsules (digital capsule) through a dynamic routing process~\cite{NIPS2017_2cad8fa4}. The routed digital capsule vectors $\textbf{v}_{j}$ are used to define the margin loss function, where $j$ denotes a class label, 0 or 1
\begin{eqnarray}
L_{j} = 
&& T_{j} \, \text{max} \Big (0,m^{+} - ||\textbf{v}_{j}|| \Big )^{2} 
+ \lambda(1-T_{j}) \, \text{max} \Big (0,||\textbf{v}_{j}||-m^{-} \Big )^{2} \;, 
\end{eqnarray} 
where a background and a signal event are labeled as $T_{0}=0$ and $T_{1}=1$, $m^+ = 0.9$, $m^- = 0.1$, and $\lambda = 0.5$.
In addition, the digital capsule vectors are connected to 3 hidden layers with increasing number of neurons from 512 to 7500 (5000) via a parallel routine to reconstruct the input image. A reconstruction loss function is defined by the sum of squared differences in pixel intensities
between the reconstructed ($\mathcal{I}^{\text{(reco)}}_i$) and the input ($\mathcal{I}^{\text{(input)}}_i$) 
\begin{eqnarray}
L_{\text{reco}} = \frac{1}{N} \sum_{i=1} ( \mathcal{I}^{(\text{reco})}_i - \mathcal{I}^{\text{(input)}}_i)^2 \;, 
\end{eqnarray} 
where the index $i$ runs from 1 to the total number of pixels in the image, and $N$ is a normalisation factor
defined by a total number of pixels in the image data times the total number of training events. The total loss function is given by
\begin{eqnarray}
L =  L_{j}  + \alpha L_{\text{reco}} \, ,
\end{eqnarray} 
where we multiply the reconstructed loss function by a scaling factor of $\alpha = 5.0 \times 10^{-4}$ to prevent it from significantly exceeding the margin loss.
The network is trained using a mini-batch size of 20. The learning rate is set to $5 \times 10^{-3}$ for $\mathcal I_{C\ell}$ and $6 \times 10^{-3}$ for $\mathcal I_{CN\ell}$. We use \texttt{weight\_decay}=$2.0 \times 10^{-7}$ in both networks.

\subsection{Graph neural networks}
\label{sec:gnn}
\begin{figure*}[t]
	\centering
	\begin{center}
		\includegraphics[width=1.\textwidth,clip]{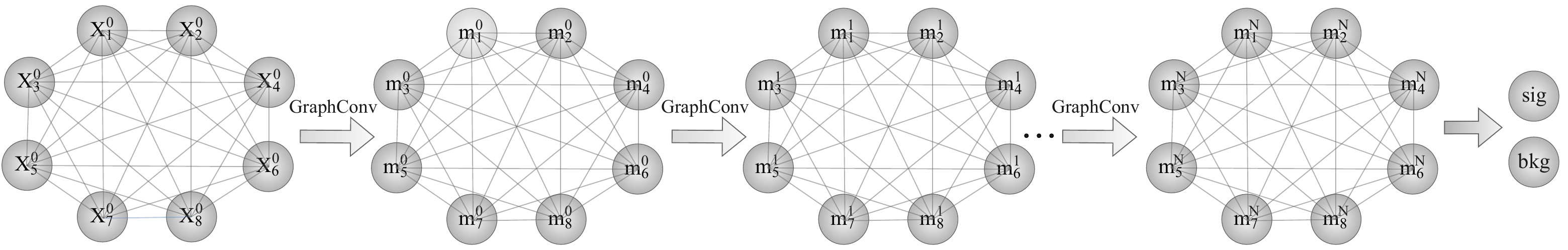}
		\caption{
			A schematic architecture of the GNN used in this paper.
		}
		\label{fig:GNN}
	\end{center}
\end{figure*}

Graph neural networks (GNN) \cite{1555942,4700287} are specialised to learn graph structured information from the data that is comprised of nodes and edges connecting them.
In this framework, each reconstructed object ($i.e.$ jets and leptons) can be represented as a node, and each node $i$ has a feature vector $\textbf{x}_i$. 
Similarly, an edge between two nodes $i$ and $j$ is represented by a vector $\textbf{e}_{ij}$ which contains an angular distance and an invariant mass of two particles.

Among various realisations of GNN models, we employ the Message Passing Neural Network (MPNN) \cite{DBLP:journals/corr/GilmerSRVD17}. In this implementation, each node $i$ has a 4-dimensional feature vector $\textbf{x}^0_i = (p_x,\ p_y,\ p_z,\ E)_i$. Each edge vector $\textbf{e}_{ij}$ contains an angular distance and an invariant mass of two particles $i$ and $j$.
A prerequisite step in the MPNN algorithm is to multiply each feature vector $\textbf{x}^0_i$ by a weight matrix $\textbf{W}^0$
\begin{eqnarray}
\textbf{m}^0_i =  \textbf{W}^0  \textbf{x}^0_i \;.
\end{eqnarray}
The feature vector evolves with a series of multiplications with nearby edge vectors and weight matrices, and its recurrent relation between layers $n$ and $n-1$ is given by
\begin{eqnarray} 
\textbf{m}_i^{n} = \sum_{j \in \mathcal{E}} \textbf{W}^n  \Big( \textbf{m}_i^{n-1}\oplus \big(\textbf{W}^{'n}  ( \textbf{m}_j^{n-1}\oplus \textbf{e}_{ij}) \big) \Big)  ,
\end{eqnarray}  
where $\textbf{W}^{n}$ and $\textbf{W}^{'n}$ denote weight matrices, and $\mathcal{E}$ denotes a set of nodes that are connected with a node $i$. This process, referred to as a graph convolution, is repeated three times. After the last graph convolution, each feature vector is concatenated in one dimension and then multiplied by a weight matrix to become a vector of length 2
\begin{eqnarray} 
\hat{\textbf{p}}_k = \textbf{W}^N \textbf{m}^{N} \; ,
\end{eqnarray}
where $k$ denotes the class label, 0 or 1 for a background or a signal, respectively. As previously mentioned, the configuration of the activation function and the loss function is the same as other networks except CapsNet. The network is trained using a mini-batch size of 20, the learning rate is set to $7.0 \times 10^{-7}$, and \texttt{weight\_decay}=$2.0 \times 10^{-4}$.

\subsection{Comparison of different networks}
\begin{figure*}[t]
	\begin{center}
		\includegraphics[width=0.68\textwidth,clip]{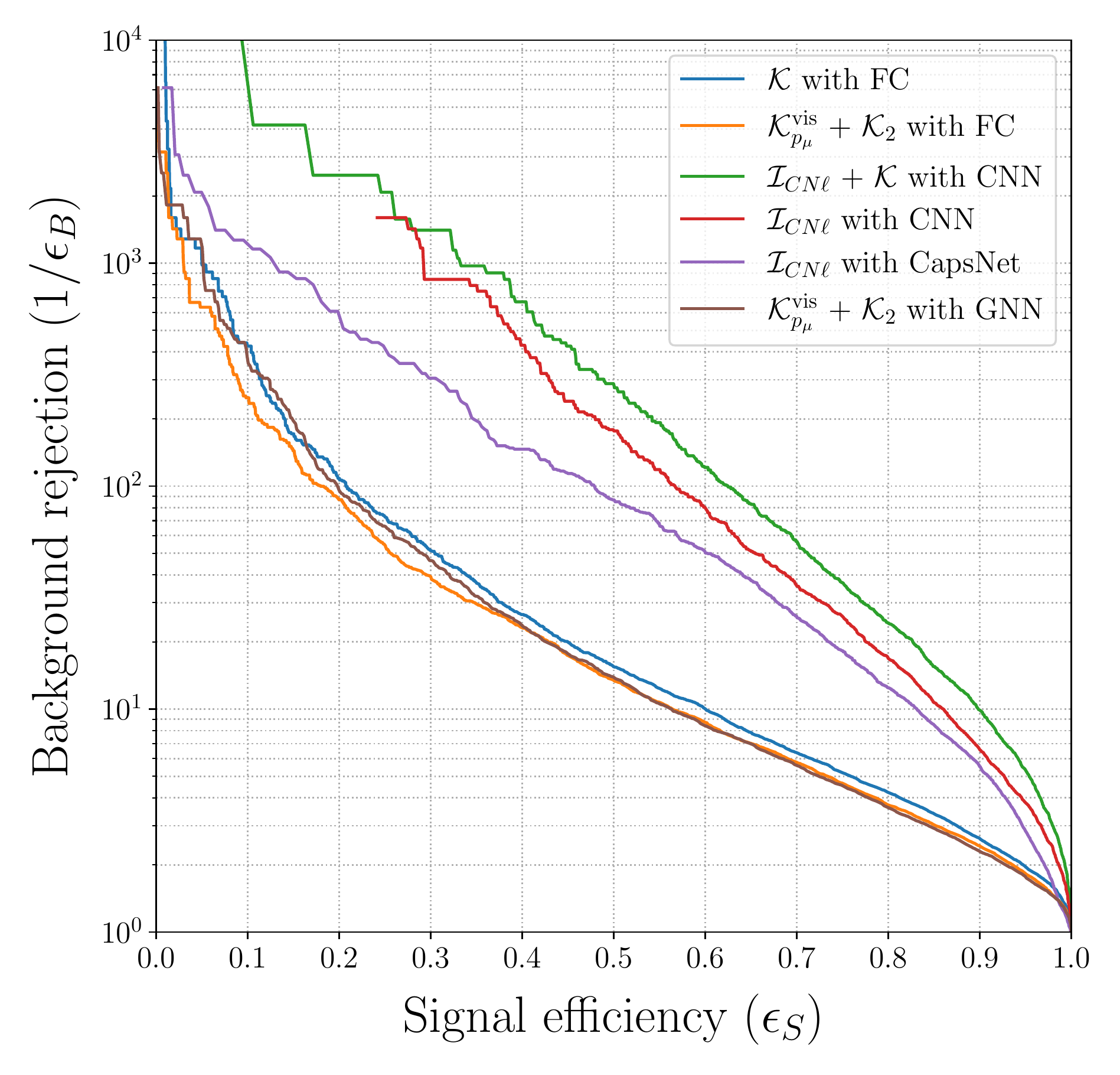}
		\caption{ROC curves for selected NN architectures evaluated on the same test sample.
			\label{fig:ML_ROC}}
	\end{center}
\end{figure*}

We present ROC
curves in \cref{fig:ML_ROC} for several architectures described above where $\epsilon_S$ and $1/\epsilon_B$ denote the signal efficiency and the background rejection, respectively.
The area under the ROC curve (AUC) is a useful measure for evaluating the performance of each architecture, where the closer the AUC is to 1, the better the network is able to classify signal and backgrounds.

We find that the CNN with $\mathcal I_{CN\ell}+\mathcal K$ has the highest AUC (0.964), followed by the CNN with $\mathcal I_{CN\ell}$ (0.951), the CapsNet with $\mathcal I_{CN\ell}$ (0.934), the FC with $\mathcal K$ (0.866), the FC with $\mathcal K_{p_{\mu}}^{\rm vis} + \mathcal K_2$ (0.852), and the GNN with $\mathcal K_{p_{\mu}}^{\rm vis} + \mathcal K_2$ (0.849). This indicates that the neural networks cannot efficiently learn important features of the signal and backgrounds when they are trained only with four-momenta information. Even when the engineered kinematic variables $\mathcal K$ are used, the results can be only mildly improved.
On the other hand, the image-based architectures generally outperform the networks that are based on the kinematic variables only. This implies that the use of image recognition techniques in the busy final states comprised of multiple jets and leptons provides a powerful tool in the search for new particles.
As also can be seen, combining the image-based CNN with the FC acting on kinematic variables provides a further gain in performance. 

Due to this comparison, we choose $\mathcal I_{CN\ell} + \mathcal K$ as data representation and a combination of a CNN acting on $\mathcal I_{CN\ell}$ and an FC acting on $\mathcal K$ in the main article.
In the main article, the additional data set $\mathcal I_{C\ell} + \mathcal K$ is used and discussed.\footnote{Comparison between CNN and CapsNet acting on $\mathcal I_{C\ell}$ (not shown in \cref{fig:ML_ROC}) show that also for this data set, CNN outperforms CapsNet.} This data set is not introduced for technical or performance reasons but because data including neutral jet images is potentially subject to corrections and uncertainties resulting from pileup effects, i.e. from the problem of separating the detector response from several particle collisions which happen simultaneously in the same bunch crossing. For charged jet images, the tracks of charged particles can be very efficiently used in order to subtract the pileup such that our projections based on $\mathcal I_{C\ell} + \mathcal K$ are conservative. 
This method has limitations for removing neutral particles, but there are on-going experimental works to cope with the problem~\cite{CMS:2020ebo,Bertolini:2014bba}, which are beyond the scope of this article. Thus, by assuming that the pileup can be controlled, the analysis based on $\mathcal I_{CN\ell} + \mathcal K$ is an optimistic (best case) scenario.

\section{Neural networks: Training and evaluation details}\label{sec:appeval}

\subsection{Choice of training data sets and comparisons}\label{sec:apptrainingset}
As described in \cref{sec:MLaspects} the data set is split into training (64\%), validation (16\%), and test (20\%) sets. For training and validation data sets, we use data comprised of background events (50\%) in which each background is added according to its respective cross section after the preselection cuts, and of signal events (50\%) for which we add equal numbers of signal events simulated with $m_S=300$~GeV to 800~GeV in steps of 50~GeV. The network is thus not trained on a specific BSM scalar mass, but instead the training aims to identify common features in signal events which distinguish the BSM events from background and apply to signal events from a broad mass window. We choose this approach because of several reasons: First, the signal contains missing energy (because we demand same-sign leptons and thus leptonic decays of a top and a $W$ boson) and a mass reconstruction is not as straight forward as in a 2-body resonance search. Second, training is expensive on computing resources, and as we show here, results from this collective training yield comparable results to re-training for a series of mass hypotheses. Third and last, training with a mixed set of different masses allows to straight forwardly apply efficiencies and bounds for all masses within the interval of $m_S=300$~GeV to 800~GeV and it gives confidence that  extrapolations above 800~GeV are feasible.\footnote{For a series of trainings at different benchmark masses, one could determine efficiencies by evaluating each trained network on each test data set with different mass signal events which can then be combined, but training costs and evaluation complexity are much higher in such an approach.}

The network is trained and validated on this ``mixed'' data set. The 15 test data sets are each specific to a single mass $m_S$ of $m_S=300$~GeV to 1000~GeV in steps of 50 GeV and comprise of (50\%) background events and (50\%) signal events simulated at mass $m_S$. To obtain projections of discovery reach and exclusion bounds at a given mass $m_S$, the fully trained network is applied to the test data set of the respective mass. The NN score is chosen as detailed in the next section, yielding a certain number of signal and background events, $S$ and $B$. We then replace $S\to \mu S$ to iteratively rescale the signal until the target significance is reached, see \cref{Eq:SigDis,Eq:SigExc}.

\begin{figure}
	\centering
	\begin{subfigure}{0.47\linewidth}
	   \includegraphics[width=\linewidth]{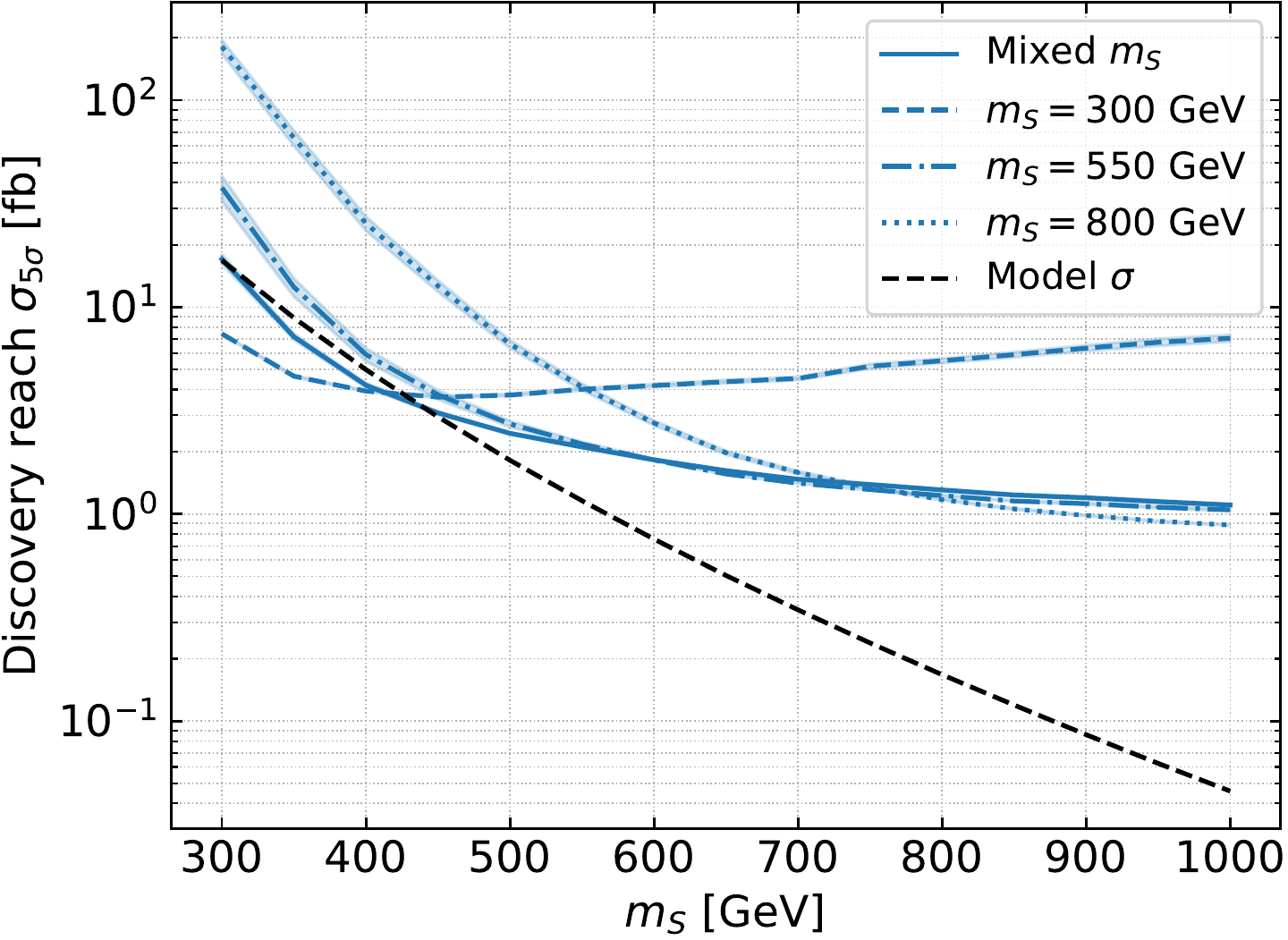}
	   \caption{Network $\mathcal K$}\label{eq:xsKfixedmass}
	\end{subfigure} \quad   
	\begin{subfigure}{0.47\linewidth}
	   \includegraphics[width=\linewidth]{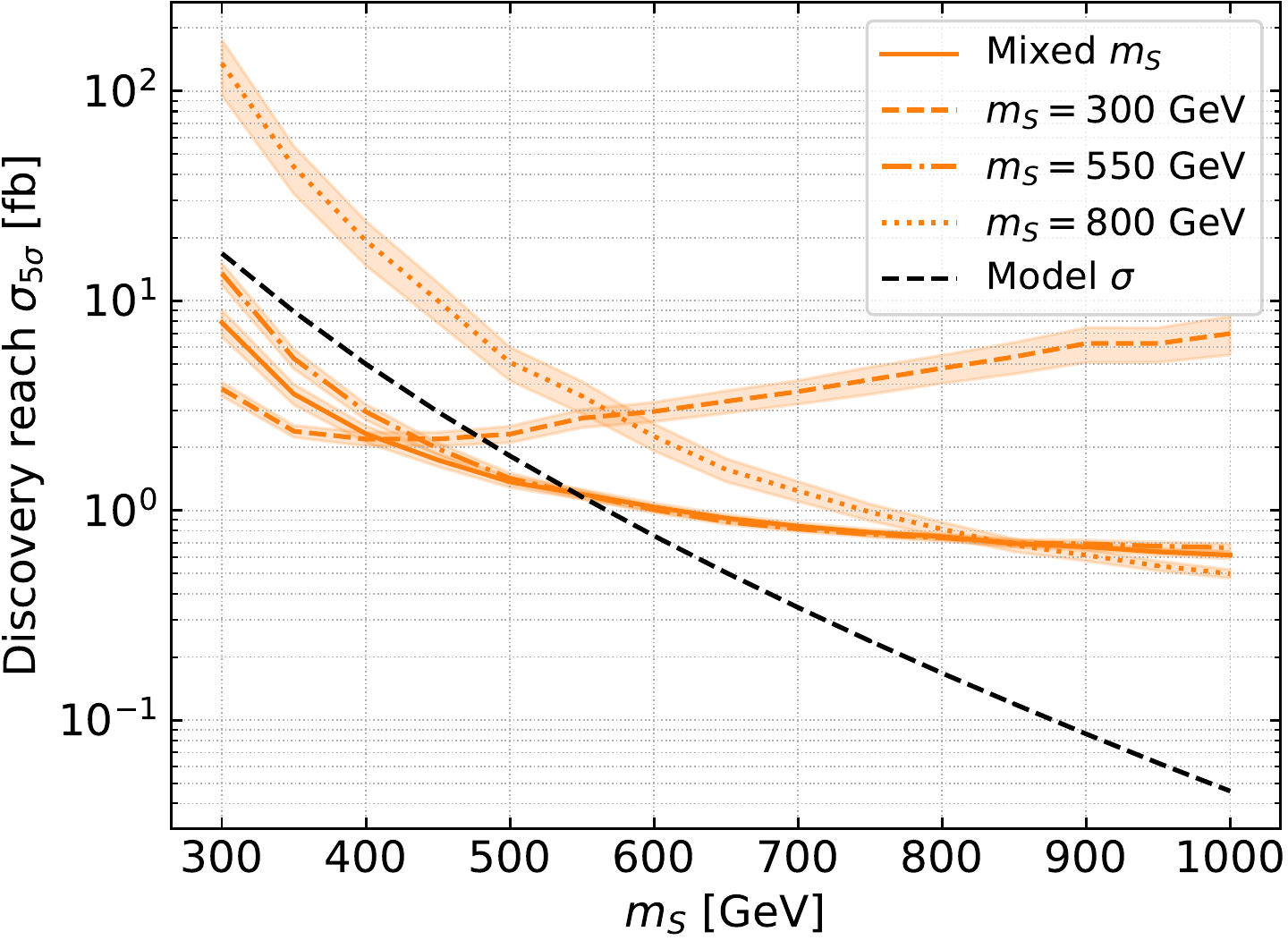}
	   \caption{Network $\mathcal I_{C\ell} + \mathcal K$}\label{eq:xsCLKfixedmass}
	\end{subfigure}\vspace{2ex}
 
	\begin{subfigure}{0.47\linewidth}
	   \includegraphics[width=\linewidth]{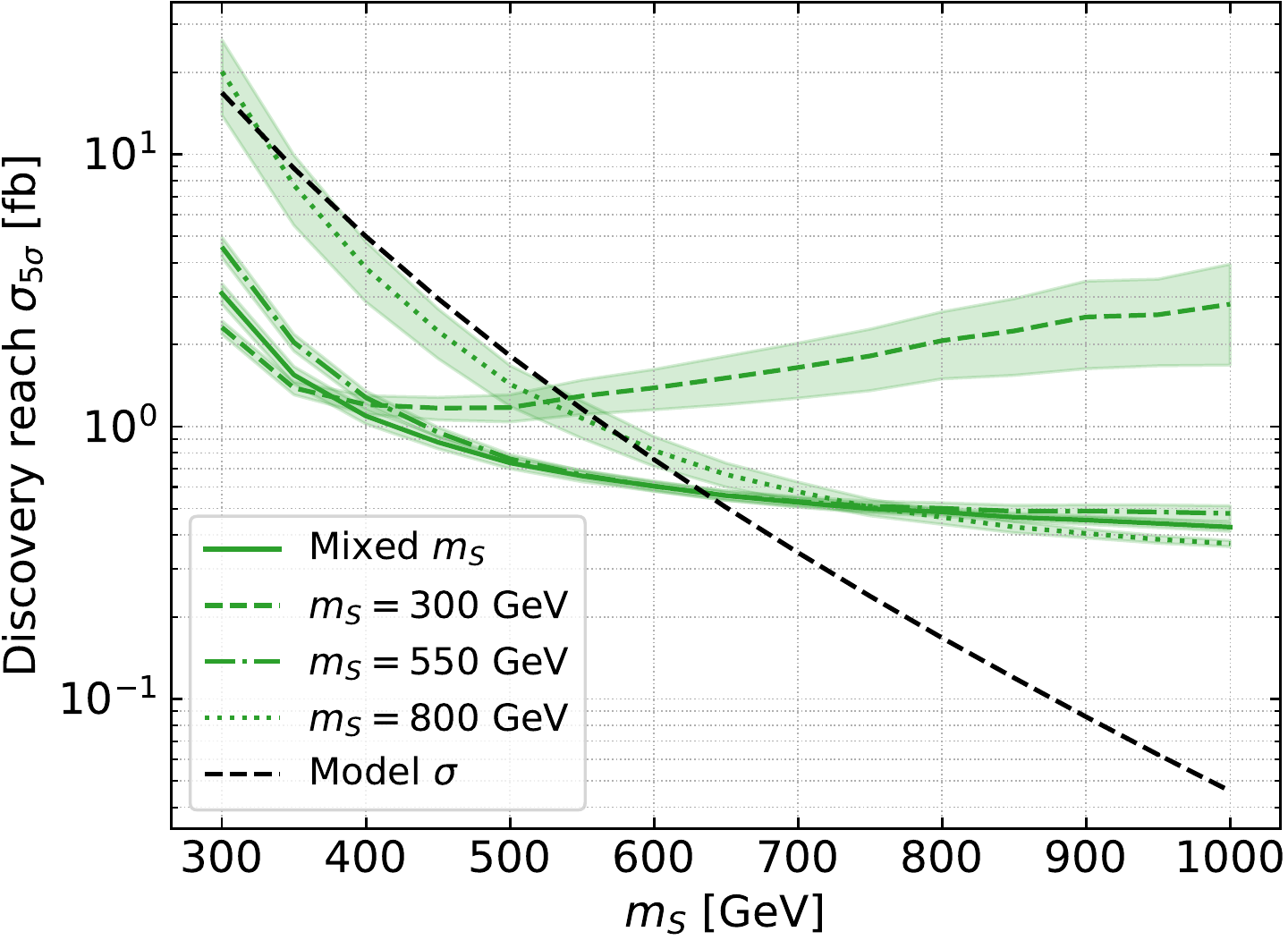}
	   \caption{Network $\mathcal I_{CN\ell} + \mathcal K$}\label{eq:xsCNLKfixedmass}
	\end{subfigure}
	\caption{Comparison of networks trained on a single fixed mass.}\label{fig:xsfixedmass}
 \end{figure}

For the main results in \cref{fig:roc,fig:xsmain,fig:xs139}, this training and evaluation procedure is independently performed 20 times for each network and each mass point, and the bands indicate the 1$\sigma$ fluctuations between the 20 different runs. For the supplementary results in \cref{fig:xsfixedmass}, the bands are based on 10 independent runs.

\bigskip

\begin{figure}
	\centering
	\begin{subfigure}{0.47\linewidth}
		\includegraphics[width=\linewidth]{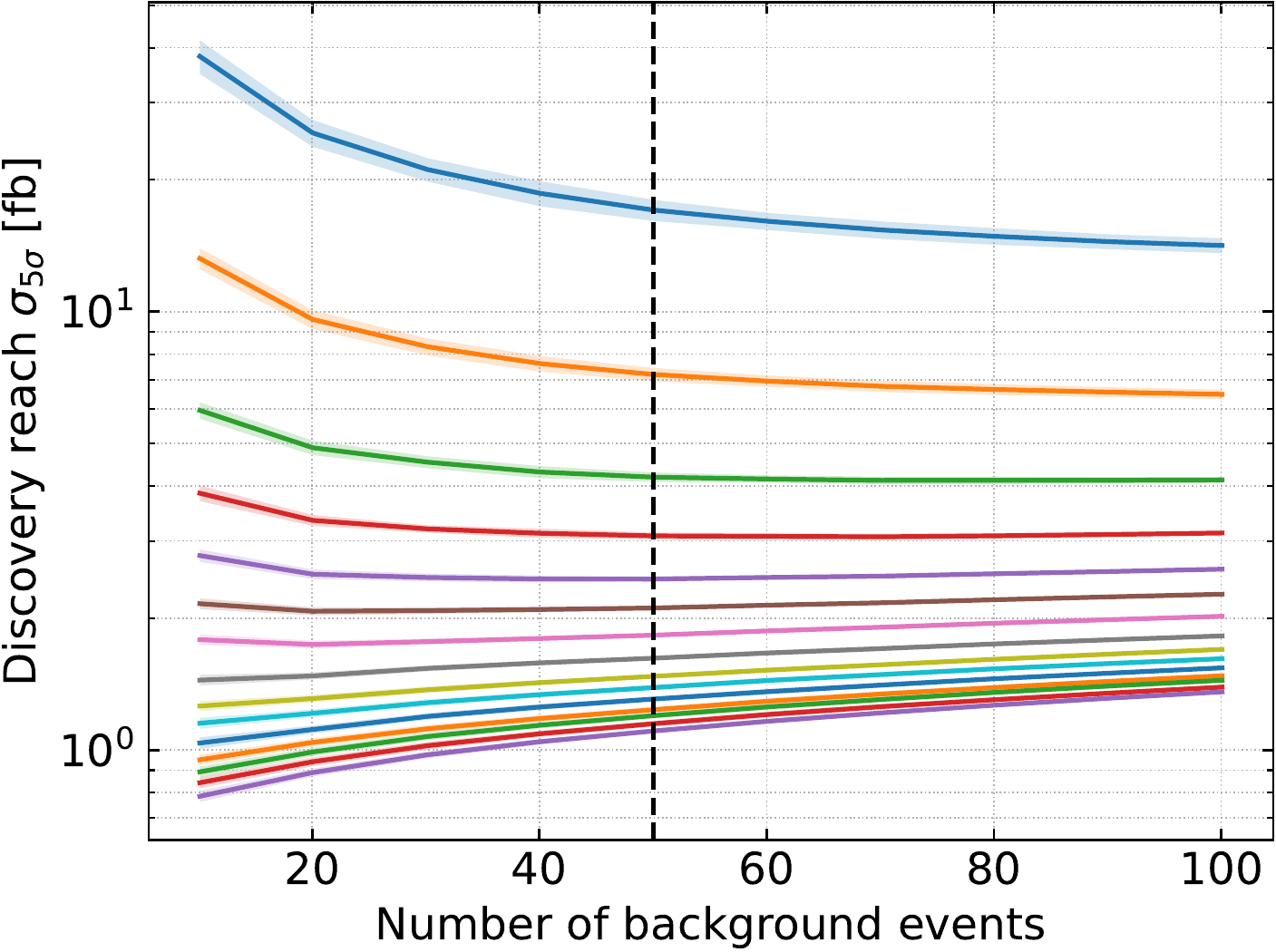}
		\caption{Network $\mathcal K$}\label{eq:xsKallmasses}
	\end{subfigure} \quad
	\begin{subfigure}{0.47\linewidth}
		\includegraphics[width=\linewidth]{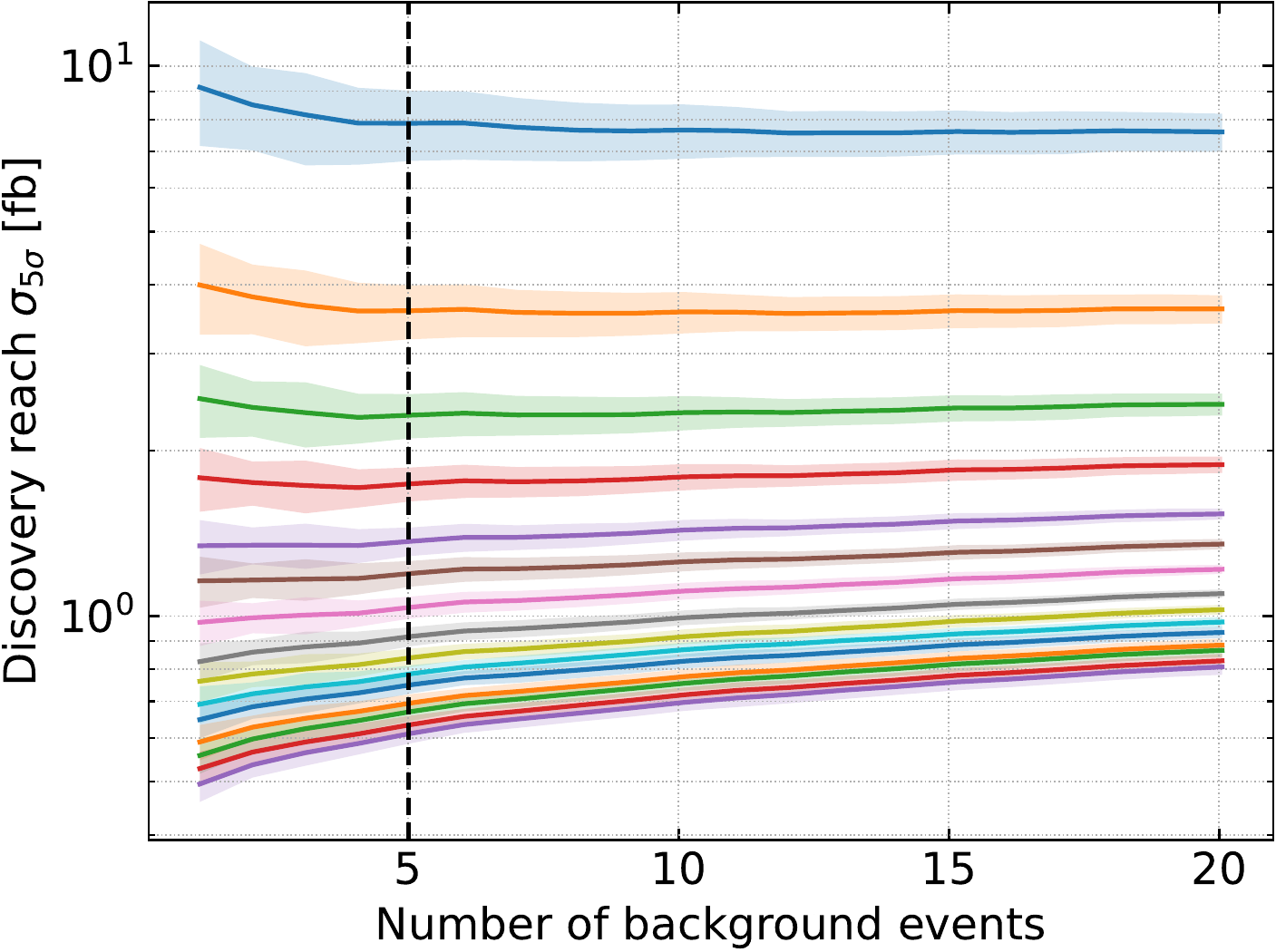}
		\caption{Network $\mathcal I_{C\ell} + \mathcal K$}\label{eq:xsCLKallmasses}
	\end{subfigure}\vspace{2ex}
	
	\begin{subfigure}{0.65\linewidth}
		\includegraphics[width=\linewidth]{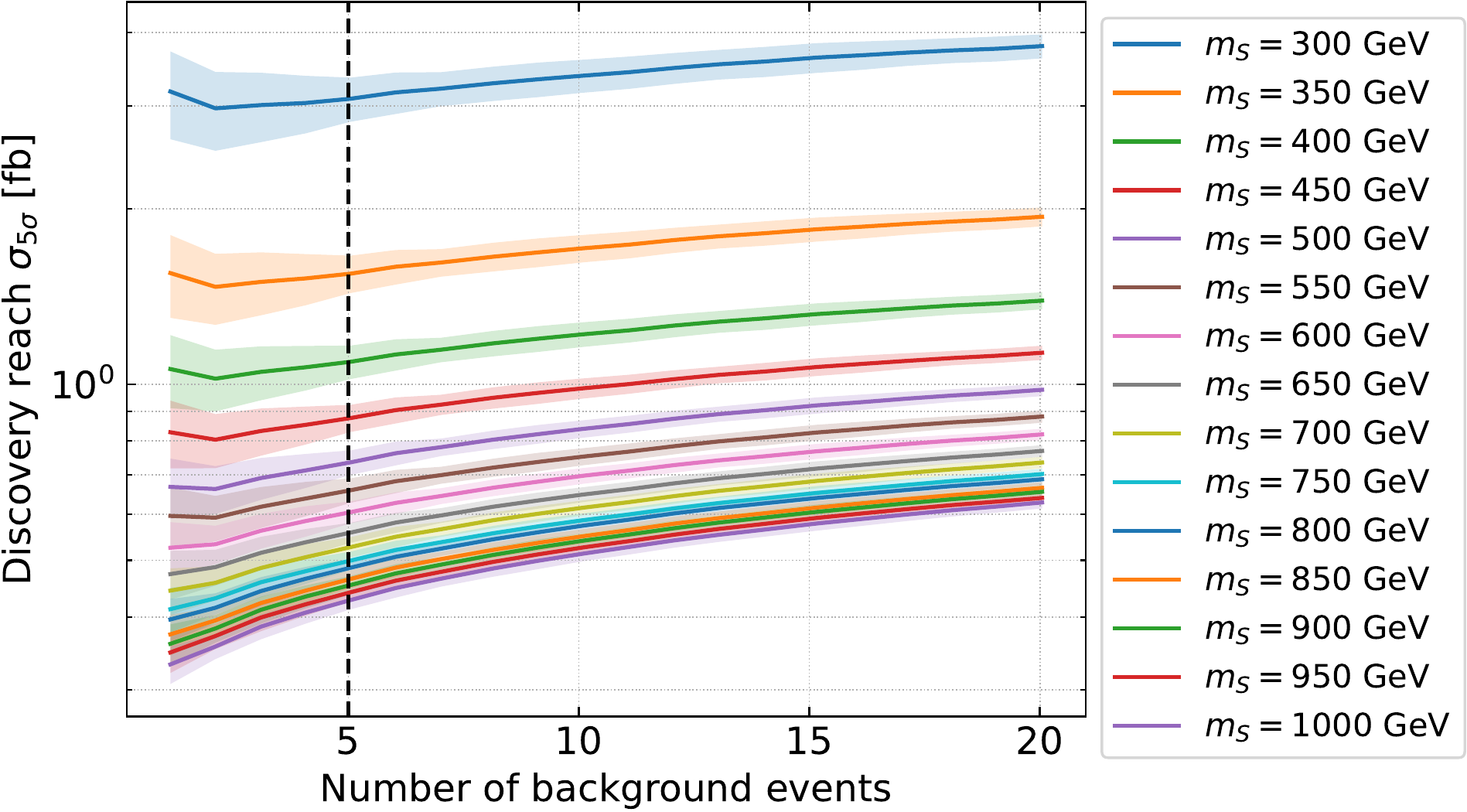}
		\caption{Network $\mathcal I_{CN\ell} + \mathcal K$}\label{eq:xsCNLKallmasses}
	\end{subfigure}
	\caption{Discovery reach as a function of number of background events, i.e.\ the NN score cut.}\label{fig:xs_allmasses}
\end{figure}

To test how well the training on a mixed data set is performing, we 
choose benchmark points at $m_S= 300$~GeV, 550 GeV and 800 GeV for which we train additional networks on a single mass. When each of these networks are evaluated  at the mass they were trained for, one would expect better performance at this mass as compared to the performance of the network trained on the mixed training set. In \cref{fig:xsmain1,fig:xsmain2}, the stars indicate the results of the mass-specific trainings. At $m_S = 300$~GeV, results from the mass specific training are indeed significantly better. However, for 550 GeV and 800 GeV, this effect is reduced or even lost, which shows that at higher masses, the training with the mixed training data set performs comparably while saving computation resources. To present some more detailed results on the mass specific training, \cref{fig:xsfixedmass} shows the discovery reach of the networks trained at a fixed mass $m_S = 300$~GeV, 550~GeV, and 800~GeV as compared to the discovery reach of the networks trained on the mixed mass data set. At $m_S=300$~GeV, the network trained on 300~GeV events performs best, but it gets quickly overtaken by the network trained on 550~GeV in the medium mass range. Only above 800~GeV  is there an advantage to training on high mass events. Even then, the improvement over the $550$~GeV network is only marginal.

Overall, the mixed data set performs well when compared to networks trained on a single mass. Only in the low mass region is there a significant difference. Therefore, a hybrid approach may be best to maximise the sensitivity of the search. Two separate networks could be trained: one for the region $m_S\leq 400$~GeV, and one for $m_S\geq 400$~GeV. The latter can either be trained on a mixed to data set or a single mass from the region of interest, e.g.\ 550~GeV.

\subsection{Cut on the neural network score}\label{sec:appcut}
When evaluating a trained network on the test set, we obtain a NN score distribution, i.e.\ each event is assigned a score between 0 (background-like) and 1 (signal-like). 
In order to derive physical results from this distribution, we need to place a cut on the NN score, below which events are discarded. 
To maximise the sensitivity of our search, the cut aught to be placed such as to maximise the signal to background ratio. 
This is typically enforced by choosing the cut to maximise the discovery significance \cref{Eq:SigDis}. 
However, we found that this method often places the cut very close to a NN score of 1, resulting in very few background events. 
Designing a search with an expected number of background events of $<1$ raises questions about the statistical soundness of the result.

To address this challenge, we plot the discovery reach as a function of the number of background events passing the NN score cut, see \cref{fig:xs_allmasses}. 
Fortunately, this reveals that the 5$\sigma$ cross section reach is not very sensitive to the number of background events. In other words: 
For the specific signature and the networks we study, decreasing the NN cut increases the number of signal and background events in such a way that there is only a mild change in discovery reach. 
We therefore fix the NN score by demanding a certain number of background events to pass the cut. 
For the networks $\mathcal K$, $\mathcal I_{C\ell}+\mathcal K$, and $\mathcal I_{CN\ell}+\mathcal K$, we require 50, 5, and 5, background events, respectively. 
We fully admit that these numbers are chosen ad hoc. 
They give sufficient statistics and at the same time do not grossly degrade the discovery potential. 

\newpage

\bibliographystyle{JHEP}
\bibliography{bibliography}

\end{document}